\documentclass[nocrop,webpdf,modern,large,namedate]{oup-authoring-template}

\graphicspath{{Figure/}}

\usepackage[mathlines, switch]{lineno}

\theoremstyle{thmstyleone}%
\theoremstyle{thmstyletwo}%
\theoremstyle{thmstylethree}%

\usepackage{bbm}
\usepackage[T1]{fontenc}
\def\code#1{\texttt{#1}}

\PassOptionsToPackage{colorlinks=true,linkcolor=blue,breaklinks}{hyperref}
\usepackage{hyperref}

\usepackage[nameinlink]{cleveref}

\crefname{figure}{Fig.}{Figs.}
\crefname{appendixfigure}{supplementary Fig.}{supplementary Figs.}
\crefname{appendixtable}{supplementary Table}{supplementary Table}
\crefname{section}{appendix}{appendix}

\begin{document}

\journaltitle{Preprint}
\copyrightyear{xxxx}
\appnotes{Paper}

\firstpage{1}


\title[\textit{CRISP}]{\textit{CRISP}: A Framework for Cryo-EM Image Segmentation and Processing with Conditional Random Field}

\author[1,$\ast$]{Szu-Chi Chung\ORCID{0000-0002-3137-6188}}
\author[1]{Po-Cheng Chou}

\authormark{Chung et al.}

\address[1]{\orgdiv{Department of Applied Mathematics}, \orgname{National Sun Yat-sen University}, \orgaddress{\street{No.70 Lien-hai Road}, \postcode{804}, \state{Kaohsiung}, \country{Taiwan}}}

\corresp{$^\ast$To whom correspondence should be addressed.}




\abstract{\textbf{Motivation:} Differentiating signals from the background in micrographs is a critical initial step for cryogenic electron microscopy (cryo-EM), yet it remains laborious due to low signal-to-noise ratio (SNR), the presence of contaminants and densely packed particles of varying sizes. Although image segmentation has recently been introduced to distinguish particles at the pixel level, the low SNR complicates the automated generation of accurate annotations for training supervised models. Moreover, platforms for systematically comparing different design choices in pipeline construction are lacking. Thus, a modular framework is essential to understand the advantages and limitations of this approach and drive further development. \\
\textbf{Results:} To address these challenges, we present a pipeline that automatically generates high-quality segmentation maps from cryo-EM data to serve as ground truth labels. Our modular framework enables the selection of various segmentation models and loss functions. We also integrate Conditional Random Fields (CRFs) with different solvers and feature sets to refine coarse predictions, thereby producing fine-grained segmentation. This flexibility facilitates optimal configurations tailored to cryo-EM datasets. When trained on a limited set of micrographs, our approach achieves over 90\% accuracy, recall, precision, Intersection over Union (IoU), and F1-score on synthetic data. Furthermore, to demonstrate our framework's efficacy in downstream analyses, we show that the particles extracted by our pipeline produce 3D density maps with higher resolution than those generated by existing particle pickers on real experimental datasets, while achieving performance comparable to that of manually curated datasets from experts. \\
\textbf{Availability:} The {\it CRISP} package is available at \href{https://github.com/phonchi/CryoParticleSegment/}{https://github.com/phonchi/CryoParticleSegment/}.
\textbf{Contact:} \href{steve2003121@gmail.com}{steve2003121@gmail.com}\\}


\maketitle




\section{Introduction}
Determining the three-dimensional (3D) structure of proteins is crucial for numerous scientific applications. This capability enables the understanding of protein functions in biological processes, the development of targeted treatments for diseases, the design of enzymes for biofuel production, and the creation of effective vaccines. Recently, cryogenic electron microscopy (cryo-EM) image processing \cite{computational_review} has revolutionized the field by significantly reducing the time required compared to traditional methods such as X-ray crystallography and nuclear magnetic resonance spectroscopy. The primary computational task in cryo-EM is to differentiate the signal from the noisy micrographs. This process aids in understanding particle distribution, detecting contaminants, extracting noise for training denoising models \cite{li2022noise}, and selecting high-quality particles from micrographs — a process known as particle picking. Particle picking is the first critical step toward achieving high-resolution protein structures. However, this process is laborious and challenging due to the low signal-to-noise ratio (SNR), the presence of contaminants, contrast variations resulting from variable ice thickness, and the aggregation of particles.

Recent developments in automated particle picking based on deep learning object detection have shown promise in overcoming these challenges \cite{wagner2019sphire,bepler2019positive,zhang2025upicker}. Nevertheless, these methods are not easily adaptable for identifying irregularly shaped objects such as ice, carbon, filaments, or membranes. Moreover, the use of bounding box outputs does not provide the pixel-level precision, which limits their application in downstream analysis. Consequently, some researchers have turned to image segmentation networks, which can distinguish particles from the background and provide visualizations of particle shapes or symmetries during early processing stages \cite{zhang2019pixer,yao2020deep,george2021cassper}. Despite these advantages, the aforementioned methods face challenges owing to the low SNR in cryo-EM images, which makes it difficult to generate the accurate pixel-level annotations necessary for training supervised models. Another limitation is that the custom codebases used for segmentation frameworks are typically not user-friendly or well-maintained, thereby raising the barrier for researchers interested in adopting these approaches. Finally, current segmentation frameworks often lack spatial regularization in the segmentation maps, which results in isolated regions or irregular boundaries.  We refer readers to \Cref{sup:back} for more information.



\begin{figure*}[!hbt]%
\centering
\includegraphics[width=1.7\columnwidth]{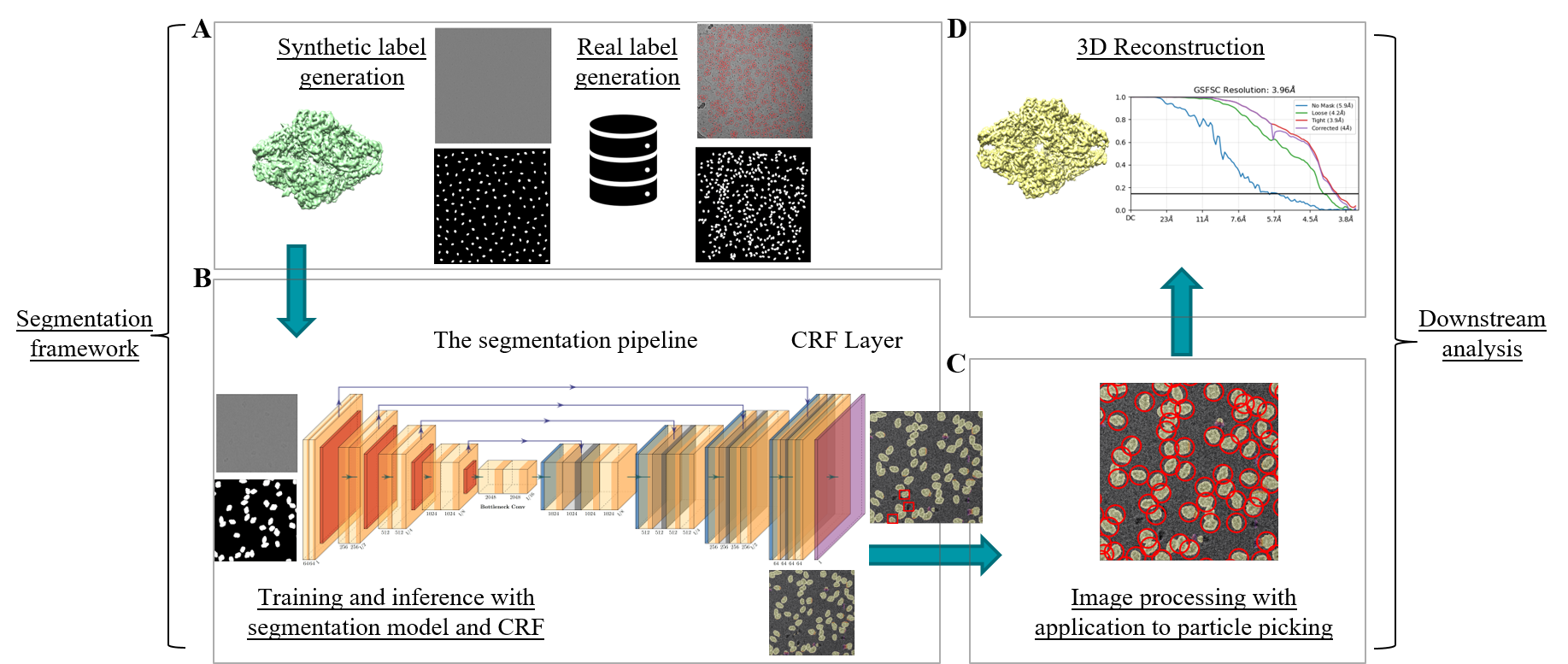}
\caption{\textit{CRISP}: The proposed framework for image segmentation and processing in cryo-EM, which can be further divided into the main segmentation and downstream analysis components. \textbf{A}: Automatic segmentation label generation process for both synthetic and real datasets. \textbf{B}: The modular segmentation pipeline that allows the user to select different architectures, loss functions, and spatial regularization methods. \textbf{C}: Downstream analysis that allows the user to automatically select the best center-finding algorithm for particle extraction. \textbf{D}: Our downstream analysis module can be seamlessly integrated with popular 3D reconstruction software.}    \label{fig:overview}
\end{figure*}

To address these issues, we propose a framework for generating segmentation maps from cryo-EM data to serve as ground truth labels for model training. To mitigate the challenges associated with custom codebases, we built our framework on top of \code{segmentation\_models.pytorch} \cite{Iakubovskii:2019}, a well-developed and flexible project that allows the selection of optimal encoders, architectures, and loss functions for constructing segmentation pipelines. Additionally, we incorporated a conditional random field (CRF) with class-discriminative features to refine weak and coarse pixel-level label predictions, resulting in sharp boundaries and fine-grained segmentation. Our results demonstrate that models trained on our labeled dataset achieve high accuracy, recall, precision, Intersection over Union (IoU), and F1-scores. Furthermore, the performance of our CRF-enhanced picker surpasses that of other state-of-the-art pickers on real datasets and is comparable to that of expert-curated particles. Notably, these models can identify particles that were not labeled in the original dataset, thereby demonstrating their generalization capability in distinguishing signals from background noise. We have organized our framework into a package, \textit{CRISP}, aimed at fostering further investigation.




\section{Methods}\label{methods}

\subsection{{\it CRISP}: The proposed image segmentation and processing framework}\label{subsec1}

Our proposed image analysis pipeline is illustrated in \Cref{fig:overview}. Notably, the framework is designed to learn at the pixel level rather than solely relying on particle shapes, thereby achieving high accuracy even on unseen micrographs. As depicted in \Cref{fig:overview}A, the label generation process efficiently creates ground truth segmentation maps from either a 3D density map or a database comprising raw micrographs, particle coordinates, and particle radius. For training a segmentation model, both the noisy micrograph and the corresponding segmentation map are provided as inputs. The segmentation pipeline is highly modular and flexible, enabling users to choose among different segmentation models, encoders, loss functions, metrics, and various CRF layers. Once the segmentation model is trained on a subset of the dataset, it can be applied to the entire dataset to differentiate signal from background at the pixel level, as shown in \Cref{fig:overview}B, with the predicted segmentation map subsequently used for downstream analysis. As a proof of concept, we include a particle picking module that employs our proposed center-finding algorithms to extract particles from the predicted segmentation map, as illustrated in \Cref{fig:overview}C. Furthermore, the extracted particle stacks are saved in the widely used \code{.star} and \code{.mrcs} formats, which can then be processed using 3D reconstruction software such as \cite{punjani2017cryosparc} to generate the final 3D density map, as shown in \Cref{fig:overview}D. Notably, the \textit{CRISP} framework is modular, allowing the downstream analysis component to be readily replaced with alternative applications, such as training denoising models \cite{li2022noise}.

\subsection{Label generation pipeline}\label{sub_chapter22}
Traditionally, obtaining a segmentation map is tedious and requires expert labeling using a carefully designed Graphical User Interface \cite{george2021cassper}. Here, we introduce a label generation pipeline that can automatically generate both synthetic and real datasets. For synthetic label generation, we download the 3D density map along with experimental information from EMPIAR \cite{iudin2023empiar}. Then, the defocus value is randomly sampled from real experiments, while the orientation is randomly sampled from $SO(3)$. Based on these parameters, we use \cite{Aspire} to generate a micrograph containing a large number of 2D projections (e.g., 100–300 particles per micrograph). Finally, we adjust the SNR to 0.005 to reflect the high noise level present in real micrographs.

For real dataset generation, inspired by \cite{zhang2019pixer}, we devise a process to automatically generate segmentation labels using real-world datasets. Firstly, we download the micrographs and the corresponding particle coordinates from EMPIAR \cite{iudin2023empiar} or CryoPPP \cite{dhakal2023large}. We then extract particles from each micrograph and use them to reconstruct the 3D density map. Next, we use the mask of the density map to generate reprojected images based on the obtained orientations obtained in 3D reconstruction. Note that we are only interested in the contour of the map; therefore, the mask is used, implying that the 3D density map need not be of high resolution and can be reconstructed using a small set of particles. Thirdly, various thresholding methods are employed to obtain binary segmented particles, which is feasible because the SNR of the reprojected image is much higher than that of the raw micrograph. Finally, we obtain the segmentation map by placing each binary segmented particle onto the micrograph according to the original coordinates.



\subsection{The modular segmentation pipeline}\label{sub_chapter23}

The segmentation pipeline begins with splitting the dataset into training, validation, and testing subsets. We then adopt a patch-based approach for training the segmentation model, which is particularly beneficial for micrographs with high particle densities \cite{zhang2025upicker}. To accelerate training, we randomly crop four patches from each training micrograph. During evaluation, the unseen micrographs are divided into overlapping patches that are fed into the segmentation network to obtain predicted segmentation patches. These patches are subsequently reconstructed using a soft weighting scheme by applying a Gaussian kernel at the interconnected boundaries, thereby producing smoother and more realistic transitions.

The pipeline is built upon \cite{Iakubovskii:2019} and allows users to experiment with different segmentation models, encoders, loss functions, and metrics. The choice of these components depends on the characteristics of the dataset, the available computational budget, and the specific challenges associated with cryo-EM. For instance, our segmentation framework supports two categories of models: (i) encoder-decoder models, which capture precise spatial hierarchies, and (ii) encoder models with simple interpolation, which are beneficial for handling multi-scale context and efficient computation, as detailed in \Cref{sup:back}. Additionally, users can select from a variety of popular convolutional neural network (CNN) families for feature extraction and choose models with different parameter sizes to balance computational time and accuracy. Furthermore, users can choose different loss functions to optimize for false positives, false negatives, or other objectives, as described in \Cref{sec:D1}. Finally, various CRF functions and solvers, as well as different postprocessing techniques for particle picking, can be selected, as detailed in the following sections.

\subsection{Spatial regularization with conditional random fields}\label{sub_chapter24}
A major obstacle in segmenting micrographs is the lack of spatial regularization in the segmentation maps, which often results in isolated regions or irregular boundaries in previous works. This problem is exacerbated by the high degree of noise and outliers present in cryo-EM data. In this context, Conditional Random Fields (CRFs) \cite{krahenbuhl2011efficient} have been introduced as a powerful method to refine weak and coarse pixel-level label predictions. Specifically, consider an image $I$ with $n$ pixels classified into $k$ classes. CRFs model the segmentation as a random field $X=\{X_1,\dots,X_n\}$, where each $X_i$ takes a value in $\{1,\dots,k\}$, and the posterior probability is modeled as follows:

\begin{equation*}
P(X=\hat{x} \mid I=\hat{I}) = \frac{1}{Z} e^{-E(\hat{x} \mid I)}
\end{equation*}
where the energy function $E(\hat{x} \mid I)$ is given by:

\begin{equation}
E(\hat{x} \mid I) = \sum_i \psi_u (\hat{x}_i \mid I) + \sum_{i<j} \psi_p (\hat{x}_i,\hat{x}_j \mid I)
\end{equation}
The unary energy components $\psi_u (\hat{x}_i \mid I)$ quantify the cost of assigning label $\hat{x}_i$ to pixel $i$, while the pairwise energy components $\psi_p (\hat{x}_i,\hat{x}_j \mid I)$ measure the cost of simultaneously assigning labels $\hat{x}_i$ and $\hat{x}_j$ to pixels $i$ and $j$. Maximizing the posterior probability $P(X \mid I)$ is equivalent to minimizing the energy function $E$.

In CRFs, the unary energies are typically obtained from a CNN \cite{zheng2015conditional,le2021regularized}, whereas the pairwise energies allow us to explicitly model interactions between pixels. In the literature, pairwise potentials are commonly modeled as weighted Gaussians:

\begin{equation}
\psi_p (\hat{x}_i,\hat{x}_j \mid I) = \mu(x_i, x_j) \sum_{m=1}^{M} w_m k_m (f_i, f_j)
\end{equation}
Here, $\mu(x_i, x_j)$ introduces a penalty for nearby pixels that are assigned different labels and can be set to the indicator function multiplied by a learnable parameter $w_0$, written as $w_0\mathbf{1}(x_i \neq x_j)$. Considering an image with feature $f_i$ that includes intensity vectors $I_i$ and positions $p_i$ at pixel $i$, the kernel $k_m$ is:

\begin{equation}
k(f_i, f_j) = w_1 e^{-\frac{|p_i - p_j|^2}{2\alpha^2} - \frac{|I_i - I_j|^2}{2\beta^2}} + w_2 e^{-\frac{|p_i - p_j|^2}{2\gamma^2}}
\end{equation}
The first term is motivated by the observation that nearby pixels with similar intensity are likely to belong to the same class, while the second term helps eliminate small, isolated regions. The parameters $w_1$ and $w_2$ are learnable. Because exact minimization of $E$ is intractable, a mean-field approximation is employed for approximate maximum posterior marginal inference \cite{zheng2015conditional}, with each step of the approximation implemented as a CNN layer for efficient training. However, as CNN architectures have grown stronger and deeper, the relative improvements provided by CRFs have diminished. Consequently, \cite{le2021regularized} introduced a regularized Frank-Wolfe algorithm that outperforms mean-field methods in early iterations, converging faster and avoiding prolonged iterations that may lead to vanishing gradients.

However, conventional CRFs rely on predefined features such as image intensity and spatial position. In our framework, we further propose the use of learning-based features similar to those in \cite{chen2022end}. Specifically, we replace the intensity feature vector $I$ in the CRF kernel with a new feature vector $g(I)$ derived from CNN feature maps — a modification we term the Class-Discriminative CRF (CD-CRF):

\begin{equation}
k(f_i, f_j) = w_1 e^{-\frac{|p_i - p_j|^2}{2\alpha^2} - \frac{|g(I_i) - g(I_j)|^2}{2\beta^2}} + w_2 e^{-\frac{|p_i - p_j|^2}{2\gamma^2}}
\end{equation}
The rationale is that CNN feature maps, particularly those from higher layers, contain more class-discriminative information than raw intensity values, which can be more usefule in cryo-EM since input intensities often fail to delineate object boundaries.

In our spatial regularization module, we employ class-dependent kernel weights to increase the number of trainable parameters, and we set $\alpha$, $\beta$, and $\gamma$ to empirical values following \cite{le2021regularized}. Moreover, we use a discriminative learning rate during fine-tuning. Finally, our framework allows the user to select different types of CRFs and CRF solvers. For more details, please refer to \Cref{sec:crf_sup}.

\subsection{Center finding and hyperparameter selection process}\label{sub_chapter25}

When the downstream analysis involves particle picking, it is essential to accurately determine the centers of particles in the predicted segmentation map. We have implemented three distinct algorithms for center finding. The first algorithm leverages a combination of morphological processing, contour extraction, filtering, and minimal enclosing box estimation to robustly detect and refine object centers. The second algorithm is based on the Crocker-Grier method \cite{crocker1996methods}. This algorithm identifies bright blobs in the segmentation map and treats those with suitable sizes as potential particles. It allows users to specify parameters such as minimum integrated brightness or particle diameter, thereby incorporating user-provided prior knowledge to address variations in particle size, mass, and symmetry. The final algorithm is based on non-maximum suppression (NMS); in this approach, the micrograph is divided into small grids, and only the maximum candidate from each grid is selected as the particle center. 

Each of these algorithms requires tuning of hyperparameters, such as grid size and the allowable overlap ratio. To address this, we implement a hyperparameter selection scheme based on the ground truth segmentation label and ground truth centers in the validation set, denoted as $\mathbf{M}_{gt}$ and $\mathbf{C}_{gt}$, respectively. The process is illustrated in \Cref{alg:hyperparam_optimization} available in the Appendix. First, we obtain the predicted centers $\mathbf{C}_{pred}$ on $\mathbf{M}_{gt}$ using different algorithms with various hyperparameter configurations. Then, we calculate the mean average precision (mAP) using $\mathbf{C}_{gt}$ and  $\mathbf{C}_{pred}$ for each configuration and select the one that achieves the highest score. Once the optimal algorithm and hyperparameters have been identified, they are applied to the entire predicted segmentation maps to obtain the final predicted centers. For more details, please refer to \Cref{sec:center_sup}.

\begin{figure*}[!hbt]%
\centering
\includegraphics[width=1.85\columnwidth]{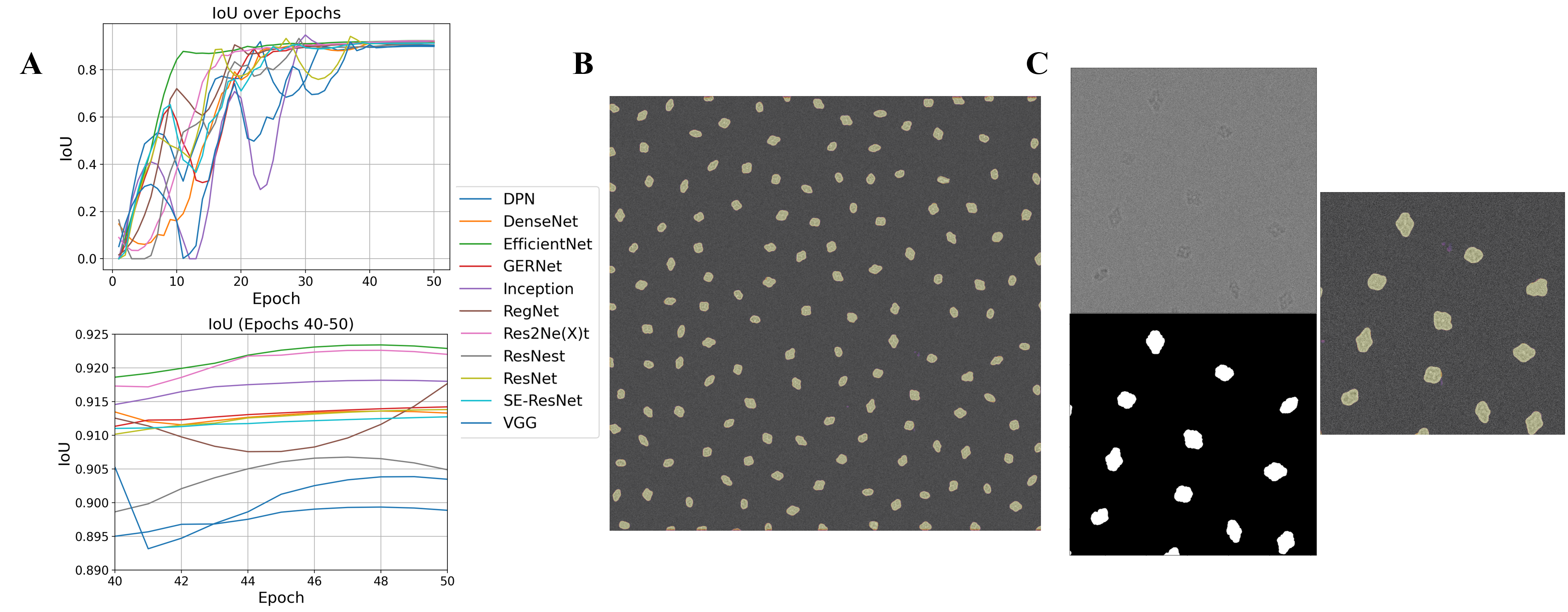}
   \caption{Results of the synthetic EMPIAR-10017 experiment. \textbf{A}: Training dynamics showing IoU versus epochs for the segmentation model using various CNN encoders available in \textit{CRISP}. The lower panel provides a zoomed-in view from epochs 40 to 50. \textbf{B}: Predicted segmentation map (highlighted in yellow) overlaid on the original micrograph, generated using U-Net++ with EfficientNet in \textit{CRISP}. \textbf{C}: The left panel displays a noisy micrograph patch along with its corresponding ground truth segmentation map, while the right panel shows a zoomed-in view of the predicted map from \textbf{B}.}
    \label{fig:simulation}
\end{figure*}

\begin{table}[hbt!]
	\centering
	\caption{Comparison of segmentation metrics on the test set of the synthetic EMPIAR-10017 dataset using various CNN encoders available in \textit{CRISP}.}
	\label{table:simulation}
\begin{tabular}{|c|c|c|c|c|}
\hline
\textbf{Architecture}                                            & \textbf{IoU}   & \textbf{Recall} & \textbf{F1 Score} & \textbf{Parameters} \\ \hline
ResNet50                                                         & 0.9138          & 0.9496           & 0.9550               & 49.0M           \\ \hline
ResNest50d                                                       & 0.9065          &  0.9484           &  0.9510               & 50.9M           \\ \hline
Res2Net50\_26w\_4s                                               &  0.9225          &  0.9554           &  0.9597               & 49.1M           \\ \hline
RegNety\_064                                                     &  0.9154          &  0.9553           &  0.9558               & 40.6M           \\ \hline
GERNet\_l                                                        &  0.9141          &  0.9538           &  0.9551               & 40.9M           \\ \hline
SE-ResNet50                                                      &  0.9127          &  0.9549           &  0.9544               & 51.5M           \\ \hline
InceptionV4                                                      &  0.9181          &  0.9534           &  0.9573               & 59.4M           \\ \hline
\textbf{Efficientnet-b5}                                         & \textbf{0.9233} & \textbf{0.9569}  & \textbf{0.9601}      & 31.9M           \\ \hline
DenseNet201                                                         &  0.9137          &  0.9525           &  0.9549               & 48.6M           \\ \hline
DPN68b                                                           &  0.8993          &  0.9359           &  0.9470               & 40.0M           \\ \hline
VGG                                                              &  0.9037          &  0.9426           &  0.9494               & 44.7M           \\ \hline
\end{tabular}
\end{table}

\section{Experiments and Results}\label{sec3}
In this study, to test the limitations of our framework, we restrict the training set and validation set to contain only 16 and 6 micrographs, respectively. In addition, U-Net++ \cite{zhou2018unet++} is selected as our segmentation model because its encoder-decoder design with skip connections is widely used in the literature. Finally, EMPIAR-10017 and EMPIAR-10081 are chosen because they are the datasets with the smallest disk size in the CryoPPP collection, allowing for quick evaluation on a personal computer. The remaining training parameters are detailed in \Cref{table:training_hyper}. For more detail information, please refer to \Cref{sup:setup}.

\subsection{\textit{CRISP} can accurately segment micrographs by leveraging its flexible architecture to achieve pixel-level precision}\label{sub_chapter31}

As a proof of concept, we applied our label generation methodologies to a synthetic dataset, using beta-galactosidase \cite{scheres2015semi} as the reference 3D map in the generation process; details are provided in \Cref{sup:setup}. To illustrate the utility of rapidly testing various architectures before selecting one well suited for segmentation, we extensively compared different CNN architectures used in the encoders and report their metrics in \Cref{table:simulation}. Note that each CNN family provides networks with different parameter sizes, and we selected the largest size that can be trained on a mainstream GPU so that they have roughly comparable parameter magnitudes. The results indicate that EfficientNet performs best in terms of IoU, recall, and F1 score on the test set, with all metrics exceeding 92\%. Recall is reported here specifically because our objective is to harvest as many particles as possible. Regarding the training dynamics, as shown in \Cref{fig:simulation}A, EfficientNet converges much faster than the other CNN architectures, requiring only 10 epochs to reach 80\% in IoU and F1 score (see \Cref{fig:simu_metric2}). Moreover, as illustrated in \Cref{fig:simulation}B and \Cref{fig:simu_ex_fig}C, the predicted segmentation map closely approximates the ground truth. When zooming into the micrograph, it is evident that all particles are accurately identified by our framework and that their overall shapes are correctly delineated, indicating pixel-level precision.

In summary, our findings demonstrate the efficacy of the proposed framework in generating supervised labels and training a highly accurate segmentation model for cryo-EM data. Furthermore, the flexibility of the framework allows for systematic comparison of different components within the pipeline —additional tests are available in \Cref{sup:more_ex}. We expect that this flexibility will enable researchers to design segmentation pipelines more systematically rather than relying solely on empirical experience. 


\begin{figure*}[hbt!]
    \centering
    \includegraphics[width=1.85\columnwidth]{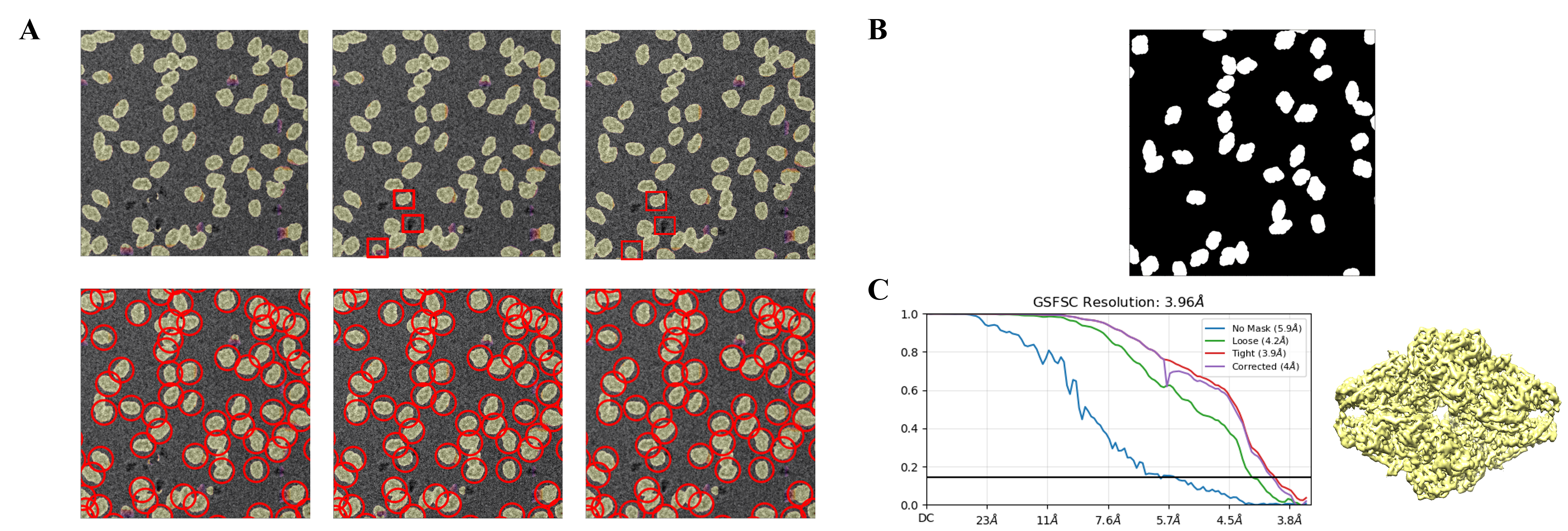}
    \caption{Results of the EMPIAR-10017 experiment. \textbf{A}: From left to right, the predicted segmentation maps overlaid on the original micrograph are shown for the original model, the model with CRF, and the model with CD-CRF, respectively. The first row displays the raw prediction maps with a red square indicating the main differences, while the second row presents the corresponding maps after applying the center finding algorithms, with a red circle indicating the selected particles. The full prediction maps are available in \Cref{fig:10017_full}. \textbf{B}: The corresponding ground truth segmentation map is shown. \textbf{C}: FSC curves and the corresponding 3D density map obtained after performing 3D reconstruction on the particles picked by \textit{CRISP}.}
    \label{fig:10017_fig}
\end{figure*}

\begin{table}[ht]
    \centering
    \caption{Performance metrics for various segmentation methods on the testing set of the EMPIAR-10017 dataset.}
    \begin{tabular}{|c|c|c|c|}
    \hline
    Method                                                         & IoU    & Recall & F1 Score \\ \hline
    U-Net++                                                        & 0.7097 & 0.8383 & 0.8302   \\ \hline
    U-Net++ with CRF                                               & 0.7116 & 0.8483 & 0.8315   \\ \hline
    \begin{tabular}[c]{@{}c@{}}U-Net++ with \\ CD-CRF\end{tabular} & \textbf{0.7129} & \textbf{0.8505} & \textbf{0.8324}   \\ \hline
    \end{tabular}
    \label{tab:crf_performance}
\end{table}

\subsection{\textit{CRISP} can restore structural signals on a real dataset by incorporating conditional random fields}\label{sub_chapter32}


We then examine the EMPIAR-10017 dataset \cite{scheres2015semi}. In this experiment, we download the curated particle coordinates and micrographs from CryoPPP and use the label generation process to produce segmentation labels. For the segmentation model, we employ the best configuration identified from the synthetic dataset, which uses EfficientNet as the encoder and Dice loss as the loss function, and train it on the real dataset. The training dynamics are shown in \Cref{fig:10017_metric}. Our segmentation model converges rapidly, requiring only 10–20 epochs to achieve reasonably good performance in terms of IoU and F1 score. However, the scores are lower than those obtained on the corresponding synthetic dataset; therefore, we visualize the predicted segmentation results in the left panel of \Cref{fig:10017_fig}A. It is evident that the predicted segmentation contains isolated regions and some of the identified particles are fragmented, indicating a degradation compared with the synthetic dataset.

We suspect that spatial regularization is particularly necessary for this more challenging dataset. To address this, we incorporate CRF and CD-CRF layers into the segmentation pipeline during model training. The results are shown in the middle and right panels of \Cref{fig:10017_fig}A, where some of the isolated regions are eliminated and fragmented particles are restored. After applying the center finding algorithm, the segmentation model with CRF and CD-CRF identifies one and two additional particles in this view, respectively. These results demonstrate that the inclusion of CRF not only improves the segmentation results at the pixel level but also enhances the downstream particle picking performance. Finally, we calculate the metrics on the test set for the three configurations and report them in \Cref{tab:crf_performance}. The results indicate that the addition of CRF improves performance, and our proposed CD-CRF outperforms the traditional CRF, with the best performing method being U-Net++ with CD-CRF.

Another factor contributing to the lower metrics compared with the synthetic dataset is that CryoPPP is a curated dataset that is not fully labeled. Consequently, it may contain false positives or false negatives due to the extremely low SNR of the micrographs. This is evident in \Cref{fig:10017_fig}, where our model identifies particles that are not present in the original labels. Therefore, the metrics alone might not be sufficient to determine which method performs best on a real dataset. To further validate the results, we perform 3D reconstruction on the particle datasets obtained from each method and compare their Fourier Shell Correlation (FSC) values. Subsequently, \Cref{alg:hyperparam_optimization} is executed, and the results are shown in \Cref{table:metrics_compare} and \Cref{fig:post_fsc_fig}. In these comparisons, the NMS method performs best and is therefore chosen for all models. From \Cref{fig:fsc_10017_unet}, it is evident that the segmentation model with CRF performs slightly better than the original model. In addition, the performance of \textit{CRISP} is comparable to that of the human-curated label dataset; the best performing method (U-Net++ with CD-CRF) achieves a resolution of 3.96 \AA{}, as depicted in \Cref{fig:10017_fig}C, compared to 4.01 \AA{} for the curated dataset. Finally, it also outperforms other state-of-the-art approaches, as shown in \Cref{table:full_compare}.

\begin{figure}[h!]
    \centering\includegraphics[width=1\columnwidth]{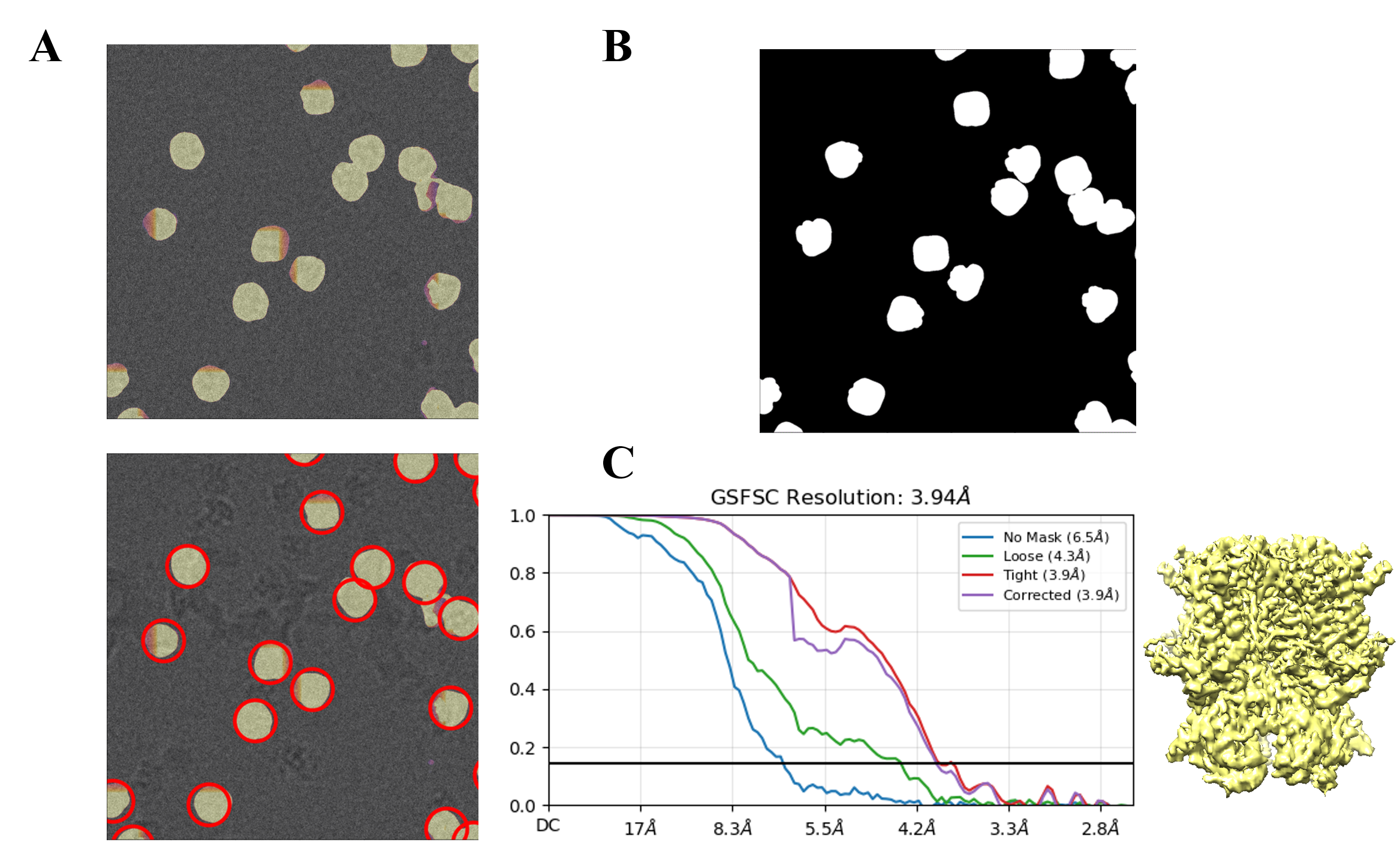}
    \caption{Results of the EMPIAR-10081 experiment. \textbf{A}: The first row shows the raw prediction map overlaid on the original micrograph, while the second row presents the corresponding map after applying the center-finding algorithms, with a red circle indicating the selected particles. The full prediction maps are available in \Cref{fig:10081_full}. \textbf{B}: The corresponding ground truth segmentation map. \textbf{C}: FSC curves and the corresponding 3D density map obtained after performing 3D reconstruction on the particles picked by \textit{CRISP}.}    
    \label{fig:10081_fig}
\end{figure}

\begin{table*}[hbt!]
	\centering
	\caption{Comparison of the number of particles obtained by various state-of-the-art methods on the EMPIAR-10081 dataset and their corresponding 3D reconstruction resolution. The compared algorithms include both object detection–based and image segmentation–based pickers (details in \Cref{sup:back}). The expert-curated dataset is also included for reference.}
	\label{table:10081_metrics}
    {\small
	\begin{tabular}{|c|c|c|c|c|c|c|c|}
 	\hline
		\textbf{EMPIAR-10081} & \textbf{CrYOLO} & \textbf{Topaz} & \textbf{Upicker} & \textbf{CASSPER} & \textbf{CryoSegNet} & \textbf{\textit{CRISP}} & \textbf{CryoPPP (Curated dataset)}\\
		\hline
		\textbf{Number of particles} & 36,821 & 37,808 & 41,164 & 27,299 & 44,819  & 42,657 & 39,352 \\
		\textbf{Resolution (\AA)} & 5.38 &  5.10 & 4.67 & 5.79 & 4.18 & \textbf{3.94} & 3.95 \\
        \hline
	\end{tabular}
   }
\end{table*}

\subsection{\textit{CRISP} achieves performance comparable to that of experts in particle picking on a challenging dataset}\label{sub_chapter33}
Next, we assessed our best model, U-Net++ \cite{zhou2018unet++} with CD-CRF, on the EMPIAR-10081 dataset \cite{lee2017structures}, which is renowned for its flexible regions and represents a challenging type of membrane protein. Notably, to test the limits of our framework, we restricted the training data to only 16 micrographs out of the 300 available in the dataset. The training dynamics are shown in \Cref{fig:10081_metric}. Our segmentation model required more time to learn the features but converged to reasonably good performance after 30 epochs in terms of IoU and F1 score. When we visualized the predicted segmentation results in \Cref{fig:10081_fig}A, some isolated regions were still present; nonetheless, the overall shape and position of the particles were well captured.

To further examine the efficacy of the predicted segmentation map in particle picking, we applied \Cref{alg:hyperparam_optimization} to the map, and the results are shown in \Cref{fig:10081_fig}A. A comparison with the ground truth labels in \Cref{fig:10081_fig}B reveals that most of the particles were correctly captured by the framework, despite some highly overlapping particles. \Cref{fig:10081_fsc} further presents the 3D reconstruction results and the FSC values obtained from different center finding algorithms. Our model achieved a resolution of 3.94 \AA{}, as depicted in \Cref{fig:10081_fig}C, demonstrating performance comparable to that of the curated dataset in CryoPPP (FSC value of 3.95 \AA{}). Furthermore, we compared the performance of \textit{CRISP} with other state-of-the-art approaches, and the results are shown in \Cref{table:10081_metrics}. Our method not only achieves the highest resolution but also harvests a reasonable number of particles. Finally, we further pushed the resolution limit by performing non-uniform refinement on both \textit{CRISP} and the CryoPPP dataset, with the results available in \Cref{fig:10081_fsc_nonuniform}. Our method achieved a resolution of 3.65 \AA{}, which is close to the published result of 3.5 \AA{}, despite using only 300 micrographs out of the entire 997 available.


\section{Discussion and Conclusion}


Distinguishing structural signals from noisy micrographs is critical for studying protein function using cryo-EM. In this study, we developed a framework to perform image segmentation on micrographs with pixel-level accuracy. To address the laborious task of manual labeling, we created a  process capable of generating segmentation maps from existing databases such as EMPIAR or CryoPPP, thereby establishing ground truth labels for training segmentation models on both synthetic and real datasets. Moreover, our framework incorporates a modular segmentation pipeline, allowing users to choose among various built-in networks, loss functions, and metrics for model training. Our experimental results suggest that, by utilizing the segmentation pipeline and the generated labels, a highly accurate segmentation model for cryo-EM micrographs can be constructed. In addition, to further refine the segmentation results, spatial regularization techniques leveraging CRF have been incorporated. Notably, a novel CRF has been proposed to improve performance based on the characteristics of cryo-EM data.

For downstream analysis, we demonstrate that the segmentation results can be directly used for particle picking by implementing a center finding algorithm. To this end, we designed a procedure to automatically select the best center finding algorithm based on the validation set. Our experimental results indicate that when using the picked particles for 3D reconstruction, our method achieves performance comparable to expert-curated particles on real datasets. Moreover, our method outperforms other state-of-the-art particle pickers on the tested datasets.

There are several topics worth exploring in the future. Firstly, preprocessing methods such as denoising or contrast adjustment may further enhance performance, as described in \cite{george2021cassper,gyawali2024cryosegnet}. Additionally, employing more powerful models, such as Transformers \cite{dhakal2024cryotransformer,zhang2025upicker}, may further improve outcomes. Regarding particle picking, utilizing our framework to pre-train a general model for automatic particle picking is a promising direction, as previous works have demonstrated that deep learning-based particle pickers can learn features that generalize well. Finally, with the predicted segmentation maps, other downstream tasks — such as training denoising models by accurately identifying noise regions \cite{li2022noise} — can be explored. We believe that current computational approaches, including particle pickers, can benefit from our modular framework. To support and stimulate further research, we have consolidated our framework into a modular package, \textit{CRISP}.




\subsection*{Funding}
This work was supported by [NSTC 112-2118-M-110-002-MY2]. 

\subsection*{Competing interests}
The authors declare no competing interests.

\bibliographystyle{natbib}
\bibliography{reference}

\clearpage
\newpage


\renewcommand{\figurename}{Supplementary Fig.}
\renewcommand{\tablename}{Supplementary Table}
\renewcommand{\theequation}{S.\arabic{equation}}

\onecolumn 
\begin{appendices}

\crefalias{figure}{appendixfigure}
\crefalias{table}{appendixtable}

\setcounter{figure}{0}  
\setcounter{table}{0} 
\setcounter{page}{1}
\setcounter{equation}{0}

\renewcommand{\thepage}{S\arabic{page}}


\section{Background}
\subsection{Background of cryo-EM and particle picking methodologies}\label{sup:back}
\begin{figure}[!bht]
	\centerline{\includegraphics[width=0.8\columnwidth]{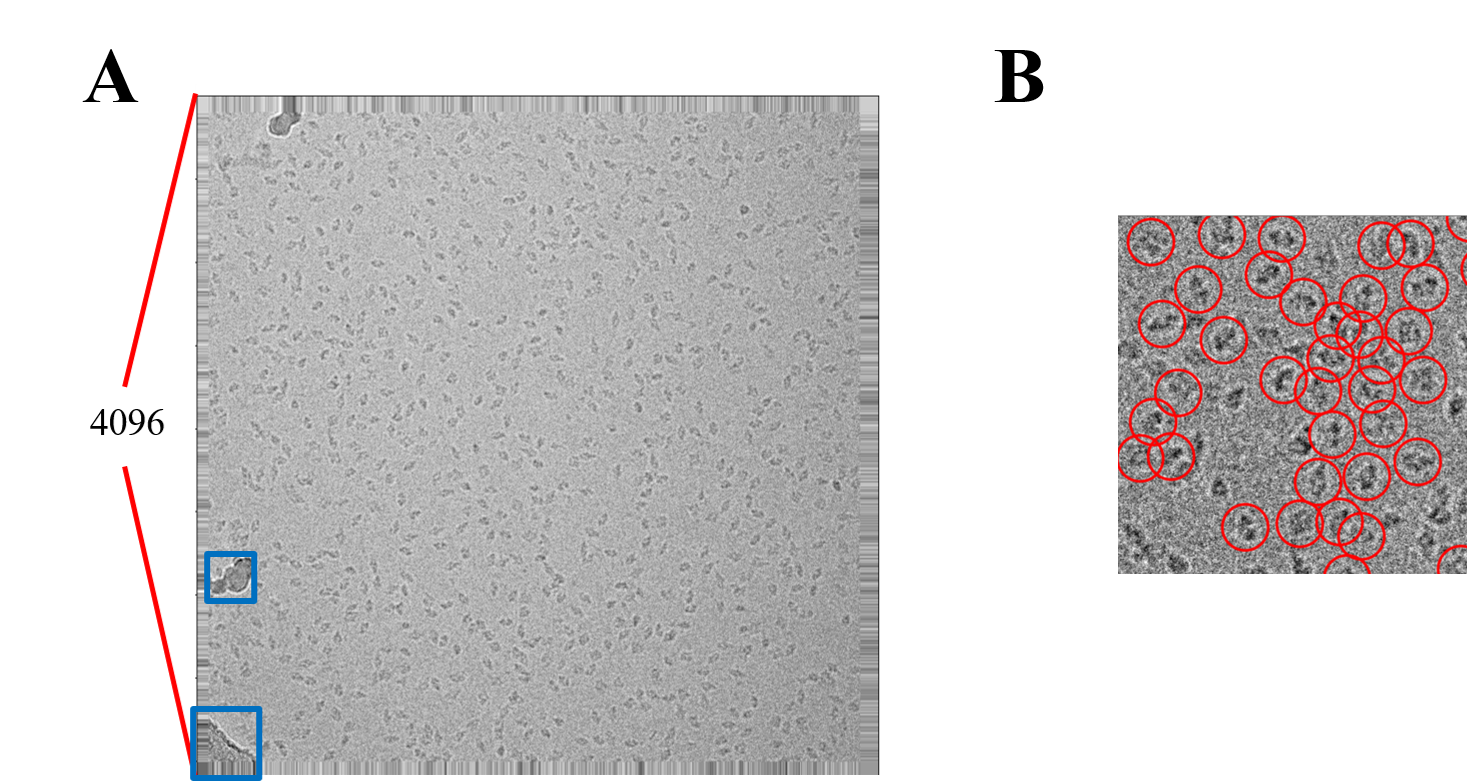}}
	\caption{\textbf{A}: A typical cryo-EM micrograph with dimensions $4{,}096 \times 4{,}096$, where the blue box indicates contaminants resembling particles. The micrograph was taken from the EMPIAR-10017 dataset and processed using \code{Topaz} \cite{s_bepler2019positive} to enhance contrast. \textbf{B}: A zoomed-in view of \textbf{A}, with a red circle indicating the particles selected from CryoPPP \cite{s_dhakal2023large}.}
    \label{fig:mic_fig}
\end{figure}

Cryo‐electron microscopy (cryo‐EM) has emerged as a transformative technology for determining the three‐dimensional structures of biological macromolecules at near‐atomic resolution without the need for crystallization. In a typical cryo‐EM experiment, biological samples are applied to grids—often coated with a thin carbon film for support—and rapidly vitrified to preserve their native state. The samples are then imaged at cryogenic temperatures under low electron doses. While this approach minimizes radiation damage and preserves protein structural information, it also produces micrographs with a very low signal-to-noise ratio (SNR) and low contrast, causing the structural details to be obscured by noise. Moreover, the micrographs contain high-contrast carbon edges resulting from the support film and ice contamination introduced during sample preparation, which can be confusing and may lead to false positives during particle picking. In some experiments, samples are used at a relatively high concentration, increasing the likelihood that particles will be densely packed or even overlap within the ice layer, especially when the ice is excessively thick. Finally, to achieve higher resolution and improve throughput, the micrographs are typically large images, resulting in high-dimensional data. All these factors make the analysis of micrographs particularly challenging, as illustrated in \Cref{fig:mic_fig}.

Consequently, a critical early step in the cryo‐EM pipeline is particle picking — the automated or semi-automated identification and extraction of individual protein particles from noisy micrographs. High-resolution three-dimensional reconstructions require the alignment and averaging of hundreds of thousands of high-quality particle images; therefore, errors during particle picking (such as missing true particles or erroneously including contaminants like ice patches, carbon edges, or malformed particles) can directly limit the accuracy of downstream reconstructions.

Early approaches to particle picking were predominantly based on template matching. In these methods, researchers manually selected a subset of particles from a few micrographs and generated one or more reference templates (often by averaging similar particles). These templates were then used to guide automated picking by scanning micrographs for regions that match the reference features, as implemented in packages such as XMIPP \cite{s_sorzano2004xmipp}, EMAN2 \cite{s_eman2}, APPION \cite{s_lander2009appion}, and RELION \cite{s_scheres2015semi}. Although template-based methods have been successfully applied for decades, their performance is highly sensitive to the inherently low SNR of cryo-EM data. Manually chosen templates may not capture the full variability in particle shape, orientation, or contrast, and can be confounded by contaminants, leading to high rates of false positives and false negatives. As a result, substantial manual post-processing and correction are often required.

In contrast, template‐free particle picking methods \cite{s_punjani2017cryosparc,s_2D_clf_relion,s_heimowitz2018apple,s_voss2009dog} are designed to identify protein particles in cryo‐EM micrographs without relying on predefined templates or manually selected examples. Instead of using reference images, these approaches leverage unsupervised computer vision techniques — such as edge detection, blob detection (e.g., using Laplacian of Gaussian filters), and local statistical measures — to automatically highlight regions that exhibit features characteristic of protein particles. By directly processing the raw image data, template‐free methods aim to detect subtle intensity variations and local maxima that differentiate particle regions from background noise. Although these methods eliminate the potential bias introduced by manually chosen templates and can be computationally efficient, they are still challenged by the inherently low SNR of cryo‐EM images, typically requiring additional post-processing or integration with other strategies to filter out false positives.

To overcome the limitations, researchers have turned to more data-driven approaches based on machine learning (ML). Early ML-based methods typically involve extracting specific image features (such as contrast, texture, or intensity statistics) from manually labeled particle regions. These features are then used to train classifiers — such as support vector machines — to distinguish particles from non-particles \cite{s_sorzano2004xmipp}. Although these approaches offer a degree of automation compared to purely template-based methods, they remain constrained by the quality and representativeness of the hand-picked training sets. Furthermore, conventional ML classifiers often struggle to generalize across datasets with varying noise characteristics and heterogeneous particle populations, highlighting the need for improved robustness and accuracy.

The advent of deep learning has revolutionized cryo‐EM particle picking by enabling models to learn complex, hierarchical features directly from raw or minimally preprocessed data. Deep learning approaches can be broadly divided into two main categories: object detection–based methods and image segmentation–based methods. In the object detection–based approach, convolutional neural network (CNN) architectures have been widely adopted. Tools such as CrYOLO \cite{s_wagner2019sphire} and Topaz \cite{s_bepler2020topaz} exemplify this approach. CrYOLO adapts the popular "You Only Look Once" (YOLO) framework to cryo‐EM images by dividing each micrograph into a grid and predicting bounding boxes for particles within each cell. Topaz, on the other hand, employs a positive-unlabeled learning paradigm, wherein a small number of annotated particles are used in conjunction with large amounts of unlabeled data. Both methods have advanced the field considerably by reducing manual intervention; however, their effectiveness is still limited by the need for extensive labeled datasets and by challenges inherent in modeling the heterogeneous appearances of particles. Recent works also leverage powerful architectures, such as Transformers, to improve the discrimination of true particles from contaminants, as demonstrated in CryoTransformer \cite{s_dhakal2024cryotransformer} and UPikcer \cite{s_zhang2025upicker}. Nevertheless, object detection–based approaches have certain limitations: they may produce off-center particle predictions, necessitating extensive translational searches for alignment in downstream analyses, and they are not easily adaptable for identifying irregularly shaped objects such as ice, carbon, filaments, or membranes. Moreover, they do not provide pixel-level precision in differentiating signal from background, which limits their application in certain scenarios \cite{s_li2022noise}.

\begin{figure}[h]
    \centering
    \includegraphics[width=0.8\columnwidth]{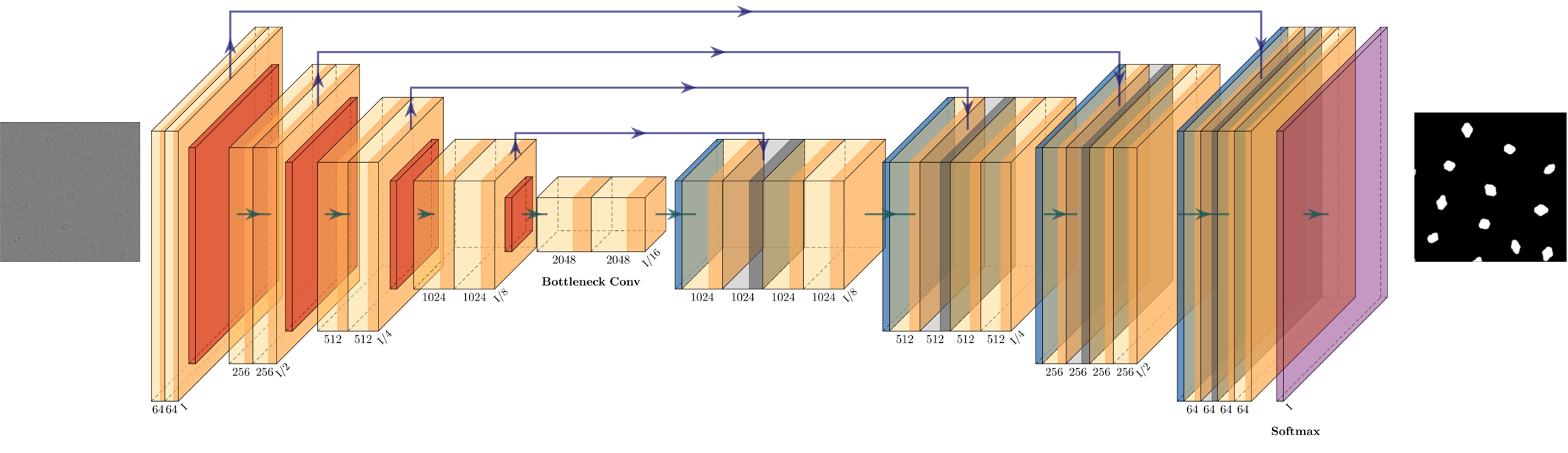}
    \caption{Schematic of the first type of image segmentation model that employs an encoder–decoder framework. In this study, we choose U-Net++ as it is one of the most widely used models in this category.}
    \label{fig:Unet}
\end{figure}

\begin{figure}[h]
    \centering
    \includegraphics[width=0.8\columnwidth]{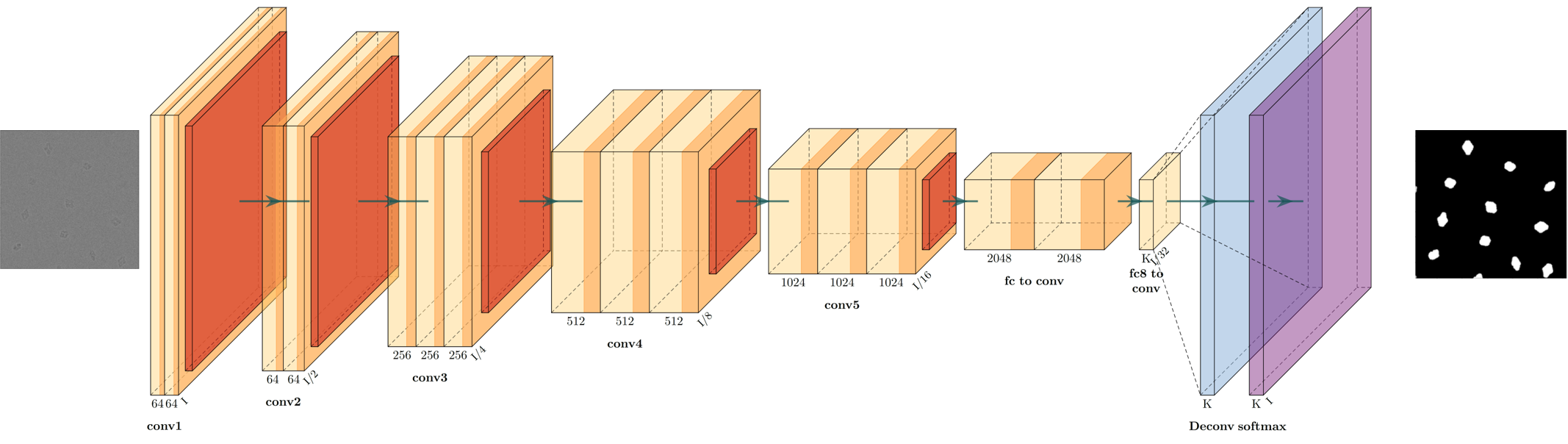}
    \caption{Schematic of the second type of image segmentation model that employs a powerful encoder paired with a simple interpolation upsampling technique. In this study, we choose DeepLabV3, as it is one of the most widely used models in this category.}
    \label{fig:Deeplab}
\end{figure}

On the other hand, recent research has explored pixel-level segmentation approaches, where the particle picking task is reformulated as an image segmentation problem. Segmentation-based methods aim to assign a class label (particle or background) to each pixel in the micrograph, thereby delineating the precise boundaries of particles. One approach follows an encoder–decoder framework exemplified by U-Net (as depicted in \Cref{fig:Unet}), which is used in previous works such as Urdnet \cite{s_ouyang2022urdnet} and CryoSegNet \cite{s_gyawali2024cryosegnet}. In this framework, a contracting encoder progressively extracts contextual features while a symmetric expanding decoder employs learned upsampling along with skip connections to recover spatial details, yielding highly precise, pixel-level segmentation maps. In contrast, other popular architectures employ Fully Convolutional Networks (FCNs) and DeepLab (as shown in \Cref{fig:Deeplab}), which use an encoder to extract robust feature representations from the input image and then rely on simple interpolation techniques or transposed convolutions to resize the feature maps back to the original resolution. This latter strategy emphasizes computational efficiency and leverages fixed upsampling schemes to reconstruct segmentation maps, as seen in PIXER \cite{s_zhang2019pixer}, PARSED \cite{s_yao2020deep}, and CASSPER \cite{s_george2021cassper}. Compared with object detection–based approaches, segmentation methods learn at the pixel level rather than relying solely on particle shapes, which may yield high accuracy even on unseen micrographs \cite{s_george2021cassper}. 

\begin{figure}[!bht]
	\centerline{\includegraphics[width=0.4\columnwidth]{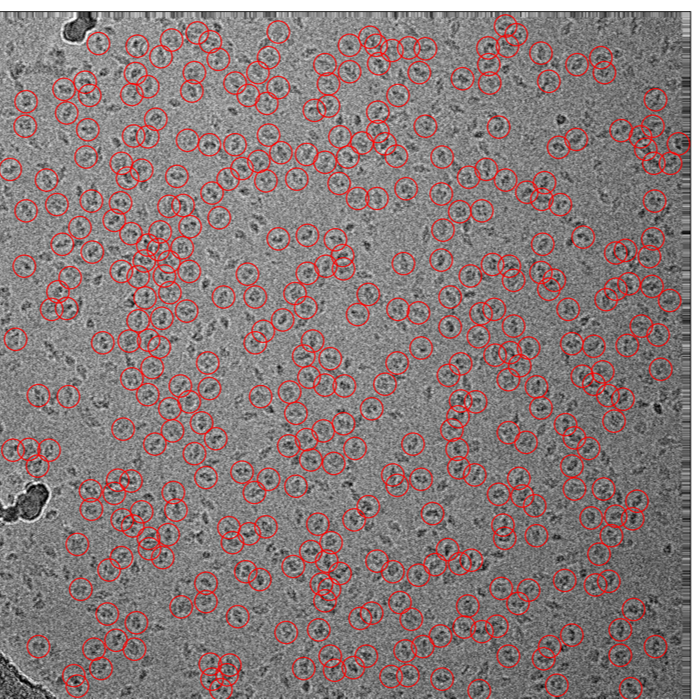}}
	\caption{A typical cryo-EM micrograph from the EMPIAR-10017 dataset processed using \code{Topaz} to enhance contrast. The red circles indicate the particles selected from CryoPPP.}
    \label{fig:obj_label}
\end{figure}

However, the image segmentation approach has some shortcomings. Firstly, it is challenging to create the accurate pixel-level annotations required for training. Secondly, the custom codebases often used for segmentation frameworks are typically not user-friendly and are difficult to modify or replace — particularly in terms of architecture and loss function — making it hard to incorporate new advances in deep learning. Finally, a significant challenge in segmentation is the lack of spatial regularization, which often leads to isolated regions or irregular boundaries, issues that are further exacerbated by the high noise levels and outliers in cryo-EM images. These challenges increase the barrier for researchers interested in adopting these approaches, despite their potential for more accurate results and broader downstream applications beyond particle picking. To address these issues, we introduce the \textit{CRISP} framework in this research.

The development of the aforementioned supervised learning methods for particle picking in cryo‐EM requires manually labeled particle datasets. The largest dataset in the cryo-EM field is the Electron Microscopy Public Image Archive (EMPIAR) \cite{s_iudin2023empiar}; however, only a small fraction of these datasets include particle labels provided by the original authors. The recently released CryoPPP \cite{s_dhakal2023large} dataset was curated from the EMPIAR repository to offer a large and diverse collection of manually labeled cryo‐EM micrographs. CryoPPP encompasses 34 representative protein datasets that cover a broad spectrum of protein sizes (ranging from approximately 77.14 kDa to 2198.78 kDa), shapes, and conformations. The collection includes micrographs that not only display ideal conditions — where particles are easily discernible — but also those featuring more challenging scenarios, such as low particle concentrations, clustered proteins, and heterogeneous views (top, side, and inclined) combined with varying defocus values and non-uniform ice distributions. Therefore, rather than using simple test datasets like Apoferritin and KLH, in this work we employ the CryoPPP dataset to evaluate our methods. Finally, since EMPIAR and CryoPPP only provide particle coordinates and radii (as shown in \Cref{fig:obj_label}), we also propose a label generation process that automatically converts this information into segmentation maps for training, as described in the main text.

\subsection{Mathmatical details of CRF}\label{sec:crf_sup}

Conditional Random Fields (CRFs) provide a rigorous probabilistic framework for structured prediction, particularly effective in tasks such as image segmentation and pixel-level labeling. In this formulation, an image is represented by a set of \( n \) pixels, where each pixel \( i \) is associated with a random variable \( X_i \) taking a value from a label set
\[
L = \{l_1, l_2, \dots, l_k\}.
\]
Given an input image \( I \), the CRF models the conditional distribution over labelings \( X = \{X_1, \dots, X_n\} \) as a Gibbs distribution:
\[
P(X \mid I) = \frac{1}{Z(I)} \exp\Bigl(-E(x \mid I)\Bigr),
\]
where \( Z(I) \) is the partition function ensuring proper normalization, and the energy \( E(x \mid I) \) is typically decomposed into unary and pairwise potentials:
\[
E(x \mid I) = \sum_{i=1}^n \psi_u(x_i) + \sum_{i<j} \psi_p(x_i, x_j).
\]
The \emph{unary potential} \( \psi_u(x_i) \) represents the cost of assigning label \( x_i \) to pixel \( i \) and is often derived from classifiers or deep CNNs. The \emph{pairwise potential} \( \psi_p(x_i, x_j) \) enforces spatial coherence by penalizing inconsistent label assignments between pairs of pixels. It is commonly expressed as:
\[
\psi_p(x_i, x_j) = \mu(x_i, x_j) \sum_{m=1}^M w_m\, k_m(f_i, f_j).
\]
Here, the index \( m \) runs over \( M \) distinct Gaussian kernels, where \( M \) denotes the number of different kernel components used to model various aspects of pixel similarity. In this formulation, \( \mu(x_i, x_j) \) is a label compatibility function. A common choice is the Potts model, defined as 
\[
\mu(x_i, x_j) = \mathbf{1}[x_i \neq x_j],
\]
which imposes a fixed penalty when the labels differ. The coefficients \( w_m \) are learnable weights associated with the \( m \)-th Gaussian kernel. The function \( k_m(f_i, f_j) \) is a Gaussian kernel applied to the feature vectors \( f_i \) and \( f_j \) of pixels \( i \) and \( j \), respectively. A typical Gaussian kernel is given by
\[
k_m(f_i, f_j) = \exp\!\left(-\frac{1}{2}(f_i - f_j)^T \Lambda_{m} (f_i - f_j)\right),
\]
where \( \Lambda_{m} \) is a symmetric positive-definite precision matrix controlling the scale and orientation of the Gaussian.

For image segmentation, the feature vector \( f_i \) often concatenates spatial coordinates \( p_i \) and color information \( I_i \). This leads to the use of two distinct kernels:
\[
\begin{aligned}
k_{\text{appearance}}(f_i, f_j) &= \exp\!\left(-\frac{\|p_i - p_j\|^2}{2\alpha^2} - \frac{\|I_i - I_j\|^2}{2\beta^2}\right), \\
k_{\text{smoothness}}(f_i, f_j) &= \exp\!\left(-\frac{\|p_i - p_j\|^2}{2\gamma^2}\right),
\end{aligned}
\]
where \( \alpha \), \( \beta \), and \( \gamma \) are hyperparameters that modulate the influence of spatial and color differences. The appearance kernel encourages pixels with similar appearance and proximity to share the same label, although its influence is generally confined to a local neighborhood. However, appearance features alone may be insufficient in regions with complex boundaries or high noise. The smoothness kernel enforces spatial coherence by promoting uniform labeling of neighboring pixels, independent of appearance. By enforcing spatial consistency, it reduces noise and refines segmentation, thereby complementing the appearance-driven kernel when the input features are not reliable.


Exact inference in fully connected CRFs is computationally prohibitive due to the quadratic number of pixel interactions. To address this, the \emph{mean field approximation} is employed to approximate the true distribution \( P(X \mid I) \); the mean field updates can be interpreted as a sequence of operations analogous to layers in a CNN \cite{s_zheng2015conditional, s_teichmann2018convolutional}. In particular, the message passing step — resembling high-dimensional filtering — can be efficiently implemented on GPUs \cite{s_krahenbuhl2011efficient, s_zheng2015conditional}. This reinterpretation not only facilitates efficient computation but also enables the integration of CRF inference into end-to-end trainable deep learning frameworks, thereby refining initial predictions (e.g., from a CNN) with improved spatial and appearance-based consistency.

However, as CNNs have grown stronger and deeper, the improvements provided by CRFs have diminished. Consequently, \cite{s_le2021regularized} introduced a regularized Frank-Wolfe algorithm that outperforms the mean field approximation in early iterations, converging faster and avoiding prolonged iterations that may lead to vanishing gradients. Regularized Frank-Wolfe optimizes a nonconvex continuous relaxation of the CRF inference problem by performing approximate conditional-gradient updates, which is equivalent to minimizing a regularized energy using the generalized Frank-Wolfe method. Moreover, they demonstrate that this approach subsumes several existing methods, including the mean field approximation as a special case, thereby enabling a unified analysis within a single framework. Within this framework, the energy gradient must be computed at each iteration. Since the pairwise potentials are Gaussian, this gradient computation can be performed efficiently (albeit approximately) again using high-dimensional filtering techniques, which can be incorporated as an efficient CNN layer \cite{s_zheng2015conditional, s_teichmann2018convolutional}. In our framework, we adopt the implementation from \cite{s_le2021regularized}.


\section{Center finding algorithms and hyperparameter selection}\label{sec:center_sup}
\subsection{The Implementation of center finding algorithms}
We have implemented three distinct center finding algorithms in our framework. In addition, we identify two important hyperparameters for tuning in each algorithm. The first hyperparameter, \(e\), controls the number of candidates in the initial search stage; a smaller value of \(e\) results in more candidates being discovered. The second hyperparameter, \(s\), governs the filtering step such that larger values of \(s\) result in more particles being filtered out.

\paragraph{Algorithm 1: \emph{Traditional Computer Vision Methods}:}
In the first algorithm, each image undergoes normalization and erosion to enhance contrast consistency and reduce noise. Normalization scales pixel intensities, followed by a two-step erosion process: first, a small kernel (of size \(\frac{\texttt{radius of expected particle}}{4}\)) is applied for noise reduction, then a second erosion is performed using a kernel of size \(\frac{\texttt{radius of expected particle}}{e}\) (with \(e \in [2,4,6]\)) to refine object boundaries and improve separation from the background. Contour detection are then performed to extract object edges. Contours are filtered by area constraints to retain only those within the range:
\[
\texttt{radius of expected particle}^s \leq \text{area} \leq 500{,}000,
\]
where \( s \in [0.6,1.0,1.4] \), ensuring the removal of irrelevant small or large regions. Next, center estimation is conducted by computing minimal enclosing circles or rectangles for each valid contour. Only enclosing circles or rectangles that satisfy predefined geometric constraints are retained, reducing the likelihood of incorrect detections.

\paragraph{Algorithm 2: \emph{Crocker-Grier Algorithm}:}
The second algorithm is based on the Crocker-Grier method \cite{s_crocker1996methods}, a classic and well-established approach for detecting and localizing blobs with pixel-level precision. This algorithm searches for local intensity maxima that serve as initial candidate locations for particles. By scanning the image with a window — often related to the expected particle size — the algorithm identifies pixels that are brighter than their neighbors, treating these pixels as the "seeds" or candidate blob centers. To reduce false positives (e.g., noise peaks), an intensity threshold is applied so that only peaks above a certain brightness are retained. Once candidate blobs are identified, the algorithm refines their centers by computing the intensity-weighted centroid within a small region around each candidate maximum. In practice, the \(x\)-coordinate of the refined center is calculated as
\[
x_{\text{center}} = \frac{\sum_{i \in \text{blob}} x_i\, I(x_i, y_i)}{\sum_{i \in \text{blob}} I(x_i, y_i)},
\]
with a similar computation for the \(y\)-coordinate and $I(x_i, y_i)$ is the intensity value at coordinate \((x,y)\). This "center of mass" approach leverages the full brightness profile of the particle. In addition, when using an expected particle size as the detection field, the algorithm calculates the center coordinates of segmented blobs and their weighted signal sums (defined as blob mass). If two or more blobs have center-to-center distances less than \(\texttt{diameter of expected particle} \times s\), where \( s \in [0.4, 0.7, 1.0] \), the blob with the smaller mass is eliminated to ensure effective picking. Finally, in our implementation, the input image is resized by a factor \(e\) (with \( e \in [0.15, 0.25, 0.35] \)) to accelerate computation and potentially collect more candidate particles.


\paragraph{Algorithm 3: \emph{Non-Maximum Suppression (NMS)}:}
The third algorithm is based on Non-Maximum Suppression (NMS) as described in \cite{s_george2021cassper}. First, each pixel in the predicted segmentation map, which represents a probability, is converted into the score of a candidate particle centered on that pixel. For a candidate particle (centered at coordinate \((x,y)\)) with a particle size of \(w \times h\), the score is computed as:
\begin{equation}
\text{score}(x, y) = \sum_{i=-\frac{w}{2}}^{\frac{w}{2}} \sum_{j=-\frac{h}{2}}^{\frac{h}{2}} W(i, j) \, P(x + i, y + j),
\end{equation}
where \(P(x, y)\) is the probability value at pixel \((x,y)\) and \(W(i, j)\) is a Gaussian kernel of size \(w \times h\) that assigns greater influence to central pixels. One benefit of using \(W(i, j)\) is that, when particles are in close proximity, the interference from neighboring particles is reduced, allowing for more precise localization. To avoid selecting overlapped particles, the micrograph is divided into small grids, and only the candidate with the maximum score from each grid is retained. The grid size is determined based on \(\texttt{diameter of expected particle} \times e\), where \( e \in [0.4, 0.5, 0.6] \). Next, a parallel local maximum search is performed to calculate the particle coordinates, with each thread covering one grid. In each iteration, the candidate is moved to the new maximum within the grid, and the threads converge to local maxima after several iterations. Finally, the remaining candidates are filtered by checking if two candidates have an overlapping region larger than \(\texttt{diameter of expected particle} \times (1-s)\), where \( s \in [0.4, 0.7, 1.0] \); if so, one candidate is removed. The remaining candidate, along with its position, is designated as the final picked particle.



\begin{algorithm}
    \caption{Hyperparameter and algorithm optimization for center-finding}
    \label{alg:hyperparam_optimization}
    \begin{algorithmic}[1]
        \Require Set of center-finding algorithms \( \mathcal{A} = \{ A_m \}_{m=1}^{M} \), set of hyperparameter configurations for each algorithm \( \mathcal{H}_m = \{ h_k \}_{k=1}^{K_m} \), ground truth segmentation map $\mathbf{M}_{gt}$, ground truth centers \( \mathbf{C}_{gt} \), bounding box size \( w, h \), IoU thresholds \( \mathcal{T} \)
        \Ensure Optimal center-finding algorithm \( A^* \) and hyperparameter configuration \( h^* \)

        \State Initialize \( \operatorname{mAP}_{\max} \gets 0 \)
        \State Initialize optimal algorithm \( A^* \gets \varnothing \)
        \State Initialize optimal hyperparameter \( h^* \gets \varnothing \)

        \For{each center-finding algorithm \( A_m \in \mathcal{A} \)}
            \For{each hyperparameter tuple \( h_k = (e_k, s_k) \in \mathcal{H}_m \)}
                \State Call the center-finding algorithm \( A_m \) with \( h_k = (e_k, s_k) \) and $\mathbf{M}_{gt}$ to obtain \( \mathbf{C}_{pred} \)
                \State Call Algorithm~\ref{alg:evaluation} with \( \mathbf{C}_{gt} \), \( \mathbf{C}_{pred} \), \( w, h \), \( \mathcal{T} \)
                \State Receive \( \operatorname{mAP}_k \) from evaluation
                \If{ \( \operatorname{mAP}_k > \operatorname{mAP}_{\max} \) }
                    \State \( \operatorname{mAP}_{\max} \gets \operatorname{mAP}_k \)
                    \State \( A^* \gets A_m \)
                    \State \( h^* \gets h_k \)
                \EndIf
            \EndFor
        \EndFor
        
        \Return \( (A^*, h^*) \)
    \end{algorithmic}
\end{algorithm}

\subsection{Hyperparameter selection}\label{sec:bounding_metric}

In this subsection, we present the metrics used in the hyperparameter search procedure. Given a set of center coordinates \( \mathbf{C} = \{(x_i, y_i)\}_{i=1}^{N} \) and a fixed bounding box size \( w, h \), each center is converted into a bounding box representation:
\[
\mathbf{B}_i = \left( x_i - \frac{w}{2},\, y_i - \frac{h}{2},\, x_i + \frac{w}{2},\, y_i + \frac{h}{2} \right)
\]
where \( \mathbf{B}_i = (x_{\min}, y_{\min}, x_{\max}, y_{\max}) \) denotes the coordinates of the bounding box. The similarity between predicted bounding boxes $\mathbf{B}_p$ and ground truth bounding boxes $\mathbf{B}_g$ is measured using the Intersection over Union (IoU), defined as:
\[
\operatorname{IoU}(\mathbf{B}_g, \mathbf{B}_p) = \frac{|\mathbf{B}_g \cap \mathbf{B}_p|}{|\mathbf{B}_g \cup \mathbf{B}_p|}
\]
where \( |\mathbf{B}_g \cap \mathbf{B}_p| \) represents the area of intersection, and \( |\mathbf{B}_g \cup \mathbf{B}_p| \) denotes the area of the union of the two bounding boxes. The IoU values are used to construct a similarity matrix:
\[
\mathbf{IoU}_{N \times M} = \left[ \operatorname{IoU}(\mathbf{B}_{g_i}, \mathbf{B}_{p_j}) \right]_{i=1, j=1}^{N, M}
\]
where \( N \) is the number of ground truth boxes and \( M \) is the number of predicted boxes. A predicted bounding box is considered a true positive (TP) if it achieves an IoU above a predefined threshold \( \tau \):
\[
\operatorname{TP} = \sum_{i=1}^{N} \sum_{j=1}^{M} \mathbf{1} \left( \operatorname{IoU}(\mathbf{B}_{g_i}, \mathbf{B}_{p_j}) \geq \tau \right)
\]
where \( \mathbf{1} \) is an indicator function that returns 1 if the condition holds and 0 otherwise. The precision \( P \) and recall \( R \) are then computed as:
\[
P = \frac{\operatorname{TP}}{M}, \quad R = \frac{\operatorname{TP}}{N}
\]
where precision measures the fraction of correct predictions among all predicted boxes, and recall quantifies the proportion of correctly detected ground truth instances.

\begin{algorithm}
    \caption{Evaluation of center-finding algorithm using mAP}
    \label{alg:evaluation}
    \begin{algorithmic}[1]
        \Require Ground truth centers \( \mathbf{C}_{gt} = \{(x_i, y_i)\}_{i=1}^{N} \), predicted centers \( \mathbf{C}_{pred} = \{(x_j, y_j)\}_{j=1}^{M} \), bounding box size \( w, h \), IoU thresholds \( \mathcal{T} \)
        \Ensure Mean Average Precision (mAP)
        
        \State Convert centers to bounding boxes:
        \For{each center \( (x, y) \) in \( \mathbf{C}_{gt} \) and \( \mathbf{C}_{pred} \)}
            \State Compute bounding box \( \mathbf{B}_i = (x - w/2, y - h/2, x + w/2, y + h/2) \)
        \EndFor

        \State Compute IoU matrix \( \mathbf{IoU}_{N \times M} \) between ground truth and predicted boxes:
        \For{each ground truth box \( \mathbf{B}_{g_i} \)}
            \For{each predicted box \( \mathbf{B}_{p_j} \)}
                \State Compute \( \operatorname{IoU}(\mathbf{B}_{g_i}, \mathbf{B}_{p_j}) = \frac{|\mathbf{B}_{g_i} \cap \mathbf{B}_{p_j}|}{|\mathbf{B}_{g_i} \cup \mathbf{B}_{p_j}|} \)
            \EndFor
        \EndFor

        \State Initialize \( \operatorname{TP} \gets 0 \), \( \operatorname{FP} \gets 0 \), \( \operatorname{FN} \gets 0 \)

        \For{each IoU threshold \( \tau \in \mathcal{T} \)}
            \State Identify true positives:
            \For{each ground truth \( \mathbf{B}_{g_i} \)}
                \If{exists \( \mathbf{B}_{p_j} \) such that \( \operatorname{IoU}(\mathbf{B}_{g_i}, \mathbf{B}_{p_j}) \geq \tau \)}
                    \State \( \operatorname{TP} \gets \operatorname{TP} + 1 \)
                \Else
                    \State \( \operatorname{FN} \gets \operatorname{FN} + 1 \)
                \EndIf
            \EndFor
            \State Compute precision and recall:
            \State \( P = \frac{\operatorname{TP}}{\operatorname{TP} + \operatorname{FP}} \), \quad \( R = \frac{\operatorname{TP}}{\operatorname{TP} + \operatorname{FN}} \)
            
            \State Compute interpolated precision:
            \For{each recall level \( r \)}
                \State \( \hat{P}(r) = \max_{\tilde{r} \geq r} P(\tilde{r}) \)
            \EndFor
            
            \State Compute Average Precision:
            \State \( AP(\tau) = \sum_{i=1}^{T} (\bar{R}_{i+1} - \bar{R}_i) \hat{P}(\bar{R}_{i+1}) \)
        \EndFor

        \State Compute Mean Average Precision:
        \State \( \operatorname{mAP} = \frac{1}{|\mathcal{T}|} \sum_{\tau \in \mathcal{T}} AP(\tau) \)
        
        \Return \( \operatorname{mAP} \)
    \end{algorithmic}
\end{algorithm}

To compute the Average Precision (AP), the precision-recall curve is interpolated as follows:
\[
\hat{P}(r) = \max_{\tilde{r} \geq r} P(\tilde{r})
\]
where precision is enforced to be non-decreasing. The AP is then obtained by integrating over recall levels:
\[
AP = \sum_{i=1}^{T} (\bar{R}_{i+1} - \bar{R}_i) \hat{P}(\bar{R}_{i+1})
\]
where \( T \) represents the number of recall levels. The mean Average Precision (mAP) is computed by averaging AP scores over multiple IoU thresholds:
\[
\operatorname{mAP} = \frac{1}{|\mathcal{T}|} \sum_{\tau \in \mathcal{T}} AP(\tau)
\]
where \( \mathcal{T} = \{0.5, 0.55, \dots, 0.95\} \) is the set of IoU thresholds. The whole procedure is summarized in \Cref{alg:evaluation}.




\begin{figure}[hbt!]
    \centering\includegraphics[width=0.8\columnwidth]{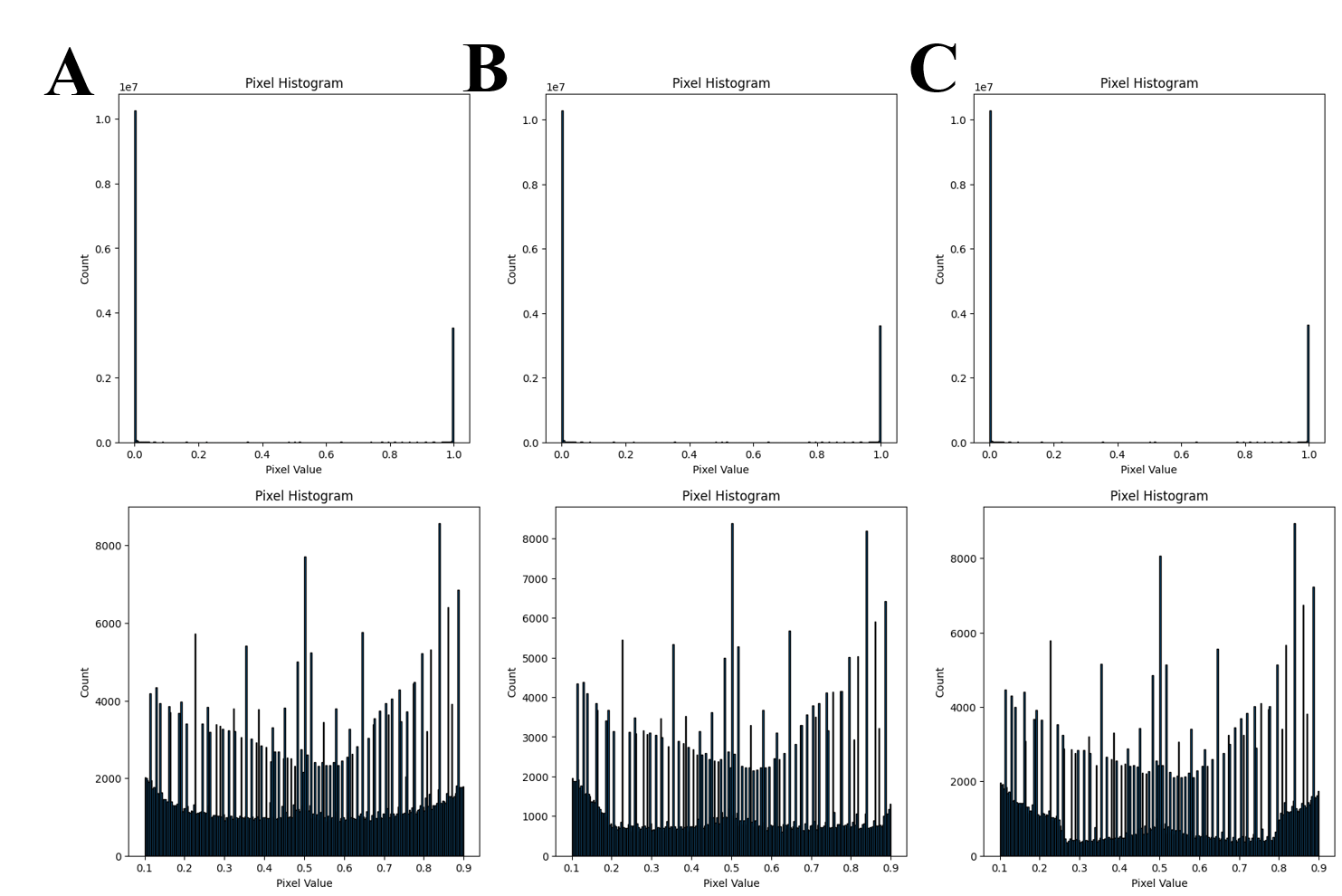}
    \caption{The upper row shows the pixel histograms of the predicted segmentation maps from the original \textbf{A} U-Net++, \textbf{B} U-Net++ with CRF, and \textbf{C} U-Net++ with CD-CRF on the EMPIAR-10017 dataset. The second row presents zoomed-in views of these three images at the central region. The pixel histograms demonstrate a progressive trend toward a bimodal distribution, highlighting the separation at the extremes of the pixel value spectrum. This pattern suggests a significant segregation of pixel intensities towards the lowest and highest possible values, illustrating the impact of the CRF.}
    \label{fig:prob_map}
\end{figure}


\section{The impact of the CRF on the predicted segmentation map}
\label{chap:appendix}

In this section, we discuss the impact of CRF inference on the predicted segmentation map. It is important to note that the output from our segmentation pipeline consists of logits, which are then converted into estimated probabilities using a softmax function. Consequently, the segmentation map is effectively a heatmap, where each pixel represents the probability that it belongs to a particle, and downstream analysis is performed based on this map.

When the center finding algorithm is applied to this map, the distribution of pixel probabilities significantly affects its accuracy. To investigate this, we plot the pixel histogram of the predicted segmentation maps obtained from the original U-Net++, U-Net++ with CRF, and U-Net++ with CD-CRF on the EMPIAR-10017 dataset, as shown in \Cref{fig:prob_map}. A detailed examination of the histogram reveals a gradual shift toward a more pronounced bimodal distribution of pixel values. In \Cref{fig:prob_map}A, two peaks appear near the extremities of the scale; however, these peaks are relatively muted, indicating a largely uniform distribution across other pixel values. After applying the CRF, as illustrated in \Cref{fig:prob_map}B, there is a noticeable increase in the counts near 0 and 1, while intermediate values decrease, resulting in more distinct peaks at the extremes. With the incorporation of the CD-CRF, the resulting histogram (shown in \Cref{fig:prob_map}C) exhibits an even more pronounced bimodal distribution, with sharp peaks at values approaching 0 and 1 and lower counts in between. This indicates that the CRF helps to better differentiate the signal from the noise — assigning probabilities close to 1 to true signal regions and close to 0 to noise — while the CD-CRF further enhances this separation, potentially improving downstream analysis.


\section{Additional experiments with the framework} \label{sup:more_ex}

\subsection{Different loss functions} \label{sec:D1}
To further illustrate the flexibility of our framework, we demonstrate how users can rapidly test various components and explore different combinations before selecting the one best suited to their needs. In this subsection, we study the impact of different loss functions on segmentation performance. The choice of loss function depends on dataset characteristics, class imbalance, and the specific challenges associated with cryo-EM. In binary segmentation, define the ground truth mask as
\[
Y = \{\, i \mid y_i = 1\,\}
\]
and the predicted mask as
\[
\hat{Y} = \{\, i \mid \hat{y}_i = 1\,\},
\]
where \(y_i\) and \(\hat{y}_i\) denote the ground truth and predicted label at pixel \(i\), respectively. The following loss functions are supported in our framework:

\begin{itemize}
    \item \textbf{Jaccard Loss:}\\[1mm]
    The Jaccard index (or Intersection over Union, IoU) is defined as
    \[
    J(Y,\hat{Y}) = \frac{|Y \cap \hat{Y}|}{|Y \cup \hat{Y}|},
    \]
    where \(|\cdot|\) denotes the cardinality (number of elements), \(Y \cap \hat{Y}\) is the set of pixels correctly predicted as positive, and \(Y \cup \hat{Y}\) is the set of pixels that are either in \(Y\) or \(\hat{Y}\). The Jaccard Loss is then given by
    \[
    \mathcal{L}_{\mathrm{Jaccard}} = 1 - J(Y,\hat{Y}).
    \]
    
    \item \textbf{Dice Loss:}\\[1mm]
    The Dice coefficient is defined as
    \[
    D(Y,\hat{Y}) = \frac{2\,|Y \cap \hat{Y}|}{|Y| + |\hat{Y}|}.
    \]
    It can also be expressed in terms of the Jaccard index:
    \[
    D(Y,\hat{Y}) = \frac{2\,J(Y,\hat{Y})}{1 + J(Y,\hat{Y})}.
    \]
    This formulation shows that the Dice coefficient places extra emphasis on the overlapping region \(Y \cap \hat{Y}\) \footnote{Consider \( f(x) = \frac{2x}{1+x} \) where \( x = J(Y, \hat{Y}) \) and \( 0 \leq x \leq 1 \). Differentiating gives \( f'(x) = \frac{2}{(1+x)^2} \), showing that \( f(x) \) is non-decreasing. Since \( f(0) = 0 \) and \( f(1) = 1 \), we have \( f(x) \geq x \) for all \( x \) in [0, 1], proving that the Dice coefficient always equals or exceeds the Jaccard index.}. The Dice Loss is then defined as
    \[
    \mathcal{L}_{\mathrm{Dice}} = 1 - D(Y,\hat{Y}).
    \]
    
    \item \textbf{Tversky Loss:}\\[1mm]
    The Tversky index generalizes the Jaccard index by weighting false negatives and false positives differently. It is defined as
    \[
    T(Y,\hat{Y}) = \frac{|Y \cap \hat{Y}|}{|Y \cap \hat{Y}| + \alpha\,|Y \setminus \hat{Y}| + \beta\,|\hat{Y} \setminus Y|},
    \]
    where
    \begin{itemize}
        \item \(Y \setminus \hat{Y}\) is the set of pixels in \(Y\) but not in \(\hat{Y}\) (false negatives), and
        \item \(\hat{Y} \setminus Y\) is the set of pixels in \(\hat{Y}\) but not in \(Y\) (false positives).
    \end{itemize}
    The Tversky Loss is then given by
    \[
    \mathcal{L}_{\mathrm{Tversky}} = 1 - T(Y,\hat{Y}).
    \]
    
    \item \textbf{Lovasz Loss:}\\[1mm]
    The Lovasz Loss is designed as a smooth surrogate for optimizing the IoU metric. The IoU is defined as
    \[
    \mathrm{IoU}(Y,\hat{Y}) = \frac{|Y \cap \hat{Y}|}{|Y \cup \hat{Y}|},
    \]
    and the Lovasz Loss approximates \(1 - \mathrm{IoU}(Y,\hat{Y})\) in a differentiable manner. For pixel-wise predictions, let \(I\) denote the set of all pixel indices and define the per-pixel Lovasz hinge loss as
    \[
    \ell_{\mathrm{Lovasz}}(\hat{y}_i,y_i) = 1 - \hat{y}_i\,y_i.
    \]
    The Lovasz Loss is then given by
    \[
    \mathcal{L}_{\mathrm{Lovasz}} = \sum_{i \in I} \ell_{\mathrm{Lovasz}}(\hat{y}_i,y_i).
    \]
    \item \textbf{Cross-Entropy Loss:}\\[1mm]
    In binary segmentation, each pixel \(i\) has a ground truth label \(y_i \in \{0,1\}\) and a predicted probability \(\hat{y}_i \in [0,1]\). Using the set notation, let
    \[
    Y = \{\, i \mid y_i = 1\,\} \quad \text{and} \quad \bar{Y} = \{\, i \mid y_i = 0\,\}.
    \]
    The Binary Cross-Entropy Loss is then defined as
    \[
    \mathcal{L}_{\mathrm{CE}} = -\sum_{i \in Y} \log(\hat{y}_i) - \sum_{i \in \bar{Y}} \log\bigl(1-\hat{y}_i\bigr).
    \]
\end{itemize}

Jaccard Loss and Dice Loss are well-suited for handling class imbalance — a common issue in cryo-EM where background pixels typically outnumber signal pixels — by directly optimizing the overlap between predicted and true segmentation masks. In contrast, Tversky Loss provides control over the balance between false positives and false negatives, making it particularly useful when the goal is to harvest more particles. Lovasz Loss, by directly optimizing the IoU metric, can ensure high-quality segmentation predictions, which may be advantageous in scenarios such as extracting noise for SNR estimation or training denoising models. Finally, Cross-Entropy Loss is valued for its ability to promote faster and more stable convergence.

We evaluated these loss functions using the best configuration identified in the main text (U-Net++ with the EfficientNet encoder) on the EMPIAR-10017 synthetic dataset, and the results are shown in \Cref{table:loss_compare}. The results indicate that Dice Loss outperforms the other loss functions in most metrics, including IoU and F1 score, which is why it is set as the default choice in this study. However, the differences are not significant when comparing different CNN encoders. Therefore, researchers can experiment with other loss functions based on their specific objectives.

\begin{table}[!hbt]
	\centering
	\caption{Comparison of segmentation metrics on the testing set of the synthetic EMPIAR-10017 dataset using various loss functions available in \textit{CRISP}. Note that the metrics are calculated according to \Cref{sup:metrics}}
	\label{table:loss_compare}
\begin{tabular}{|c|c|c|c|c|c|}
\hline
             & IoU   & Precision & Recall & Accuracy & F1   Score \\ \hline
Dice         &  \textbf{0.9233} &  \textbf{0.9633}     &  0.9569  &  \textbf{0.9953}    &  \textbf{0.9601}      \\ \hline
Jaccard      &  0.9213 &  0.9594     &  0.9587  &  0.9951    &  0.9591      \\ \hline
Lovasz       &  0.9209 &  0.9588     &  0.9587  &  0.9951    &  0.9588      \\ \hline
Tversky      &  0.9154 &  0.9509     &  \textbf{0.9608}  &  0.9947    &  0.9558      \\ \hline
Cross-Entropy &  0.9229 &  0.9620     &  0.9577  &  0.9952    &  0.9599      \\ \hline
\end{tabular}
\end{table}

\subsection{Different CRF solver}

\begin{table*}[!hbt]
	\centering
	\caption{Performance metrics for different segmentation methods on the testing set of the EMPIAR-10017 dataset. Here, MF and FW denote the mean-field and regularized Frank-Wolfe CRF solvers, respectively.}
	\label{table:metrics_crf_solver}
\begin{tabular}{|c|ccc|}
\hline
                                    & \multicolumn{3}{c|}{Unet}                                                                       \\ \hline
                                    & \multicolumn{1}{c|}{IoU   (MF/FW)} & \multicolumn{1}{c|}{Recall   (MF/FW)} & F1 Score   (MF/FW) \\ \hline
(1)   U-Net++              & \multicolumn{1}{c|}{0.7097}         & \multicolumn{1}{c|}{0.8383}            &  0.8302              \\ \hline
(2)   U-Net++ with CRF               & \multicolumn{1}{c|}{0.7117/0.7116}   & \multicolumn{1}{c|}{0.8463/0.8483}      &  0.8316/0.8315        \\ \hline
(3)   U-Net++ with CD-CRF     & \multicolumn{1}{c|}{\textbf{0.7129}/\textbf{0.7129}}   & \multicolumn{1}{c|}{0.8481/\textbf{0.8505}}      &  \textbf{0.8324}/\textbf{0.8324}        \\ \hline
                                    & \multicolumn{3}{c|}{DeepLabV3}                                                                    \\ \hline
(1)   DeepLabV3           & \multicolumn{1}{c|}{0.6918}         & \multicolumn{1}{c|}{0.8087}            &  0.8178              \\ \hline
(2)   DeepLabV3   with CRF          & \multicolumn{1}{c|}{0.6993/0.6995}   & \multicolumn{1}{c|}{0.8263/0.8268}      &  0.8231/ 0.8232        \\ \hline
(3)   DeepLabV3   with CD-CRF & \multicolumn{1}{c|}{\textbf{0.6996}/0.6995}   & \multicolumn{1}{c|}{0.8273/\textbf{0.8276}}      &  \textbf{0.8232}/ \textbf{0.8232}        \\ \hline
\end{tabular}
\end{table*}

In this subsection, we compare the impact of different CRF solvers and also test the performance of both CRF and CD-CRF on an alternative architecture, DeepLabV3. Our framework directly supports the classic mean-field solver as well as the state-of-the-art regularized Frank-Wolfe method, as discussed in \Cref{sec:crf_sup}, which potentially offers improved convergence performance. As shown in \Cref{table:metrics_crf_solver}, the regularized Frank-Wolfe solver achieves comparable performance to the mean-field approach, with a slight improvement in the Recall metric. This indicates its potential to harvest more particles during downstream particle picking, which is why it is the default choice in this study. Furthermore, our proposed CD-CRF consistently outperforms other methods when integrated with the DeepLabV3 architecture, demonstrating its robustness and effectiveness in improving segmentation results.

\begin{table*}[!hbt]
	\centering
	\caption{Metrics for all center-finding algorithms available in \textit{CRISP} on the validation sets of EMPIAR-10017 and EMPIAR-10081. The metrics are calculated according to \Cref{sec:bounding_metric}.}
	\label{table:metrics_compare}
\begin{tabular}{|c|cccc|}
\hline
                                 & \multicolumn{1}{c|}{Precision} & \multicolumn{1}{c|}{Recall} & \multicolumn{1}{c|}{mAP}   & F1 score  \\ \hline
                                 & \multicolumn{4}{c|}{EMPIAR-10017}                                                                       \\ \hline
Morphology   and contour finding & \multicolumn{1}{c|}{0.9420}     & \multicolumn{1}{c|}{0.9223}  & \multicolumn{1}{c|}{0.7829} &  0.9320 \\ \hline
Crocker-Grier   algorithm        & \multicolumn{1}{c|}{\textbf{0.9997}}     & \multicolumn{1}{c|}{0.9012}  & \multicolumn{1}{c|}{0.7588} &  0.9479 \\ \hline
Non-Maximum Suppression            & \multicolumn{1}{c|}{0.9881}     & \multicolumn{1}{c|}{\textbf{0.9736}}  & \multicolumn{1}{c|}{\textbf{0.8010}} &  \textbf{0.9808} \\ \hline
                                 & \multicolumn{4}{c|}{EMPIAR-10081}                                                                    \\ \hline
Morphology   and contour finding & \multicolumn{1}{c|}{0.9894}     & \multicolumn{1}{c|}{0.9538}  & \multicolumn{1}{c|}{0.8426} &  0.9713 \\ \hline
Crocker-Grier   algorithm        & \multicolumn{1}{c|}{\textbf{1.0000}}    & \multicolumn{1}{c|}{0.9809}  & \multicolumn{1}{c|}{\textbf{0.8890}} &  0.9904 \\ \hline
Non-Maximum Suppression            & \multicolumn{1}{c|}{\textbf{1.0000}}    & \multicolumn{1}{c|}{\textbf{0.9901}}  & \multicolumn{1}{c|}{0.8622} &  \textbf{0.9950} \\ \hline
\end{tabular}
\end{table*}

\begin{figure}[!bht]
	\centerline{\includegraphics[width=0.4\columnwidth]{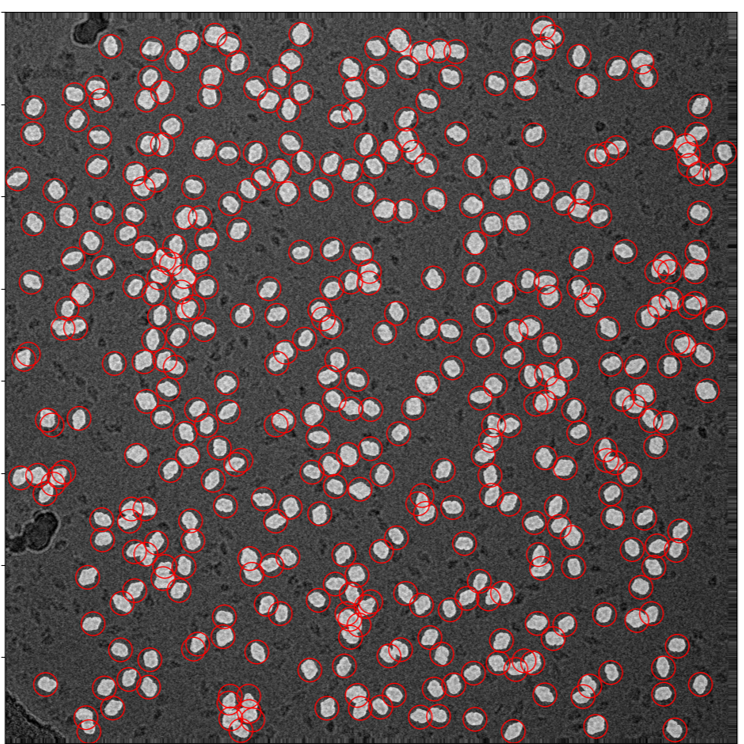}}
	\caption{A typical cryo-EM micrograph from the EMPIAR-10017 validation set overlaid with its ground truth segmentation map. The red circles indicate the particles selected by the center-finding algorithm using Non-Maximum Suppression.}
    \label{fig:nms}
\end{figure}

\subsection{Different Center Finding Algorithms}
In this subsection, we present the results of applying different center finding algorithms to the predicted segmentation map. For the segmentation model, we use the best configuration from the main text, which employs U-Net++ with an EfficientNet encoder and Dice loss as the loss function. First, we generate the predicted segmentation maps for the EMPIAR-10017 and EMPIAR-10081 datasets. Next, we apply the three center finding algorithms with different \(e\) and \(s\) parameters, as described in \Cref{sec:center_sup}, and select the optimal hyperparameters for each algorithm based on the mAP metric on the validation set. Finally, we report the precision, recall, and F1 score at an IoU threshold of 0.5, as well as the overall mAP, in \Cref{table:metrics_compare}.

\begin{figure}[h]
    \centering
    \includegraphics[width=0.9\columnwidth]{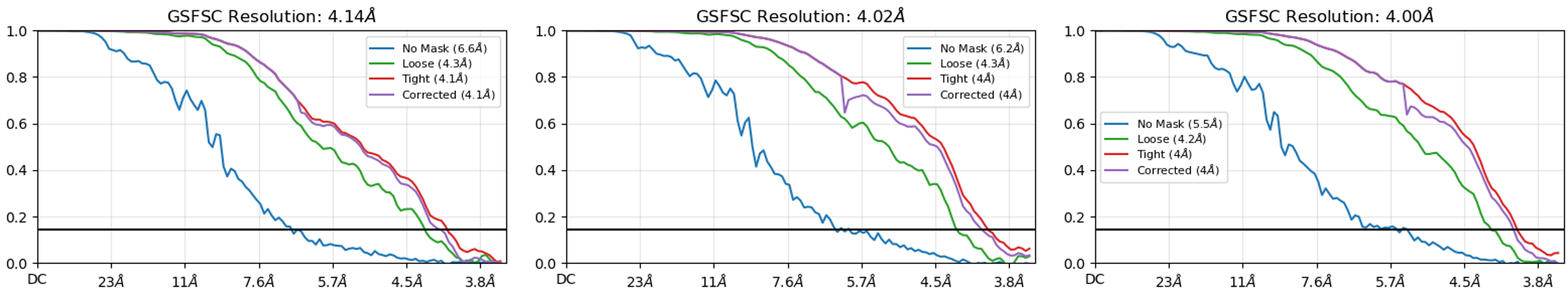}
    \caption{Visualization of the FSC curves obtained from 3D reconstruction using particles selected by different center-finding algorithms on the EMPIAR-10017 dataset using UNet++. From left to right, the panels correspond to morphology and contour finding, the Crocker-Grier algorithm, and Non-Maximum Suppression.}
    \label{fig:post_fsc_fig}
\end{figure}

\begin{figure}[h]
    \centering
    \includegraphics[width=0.9\columnwidth]{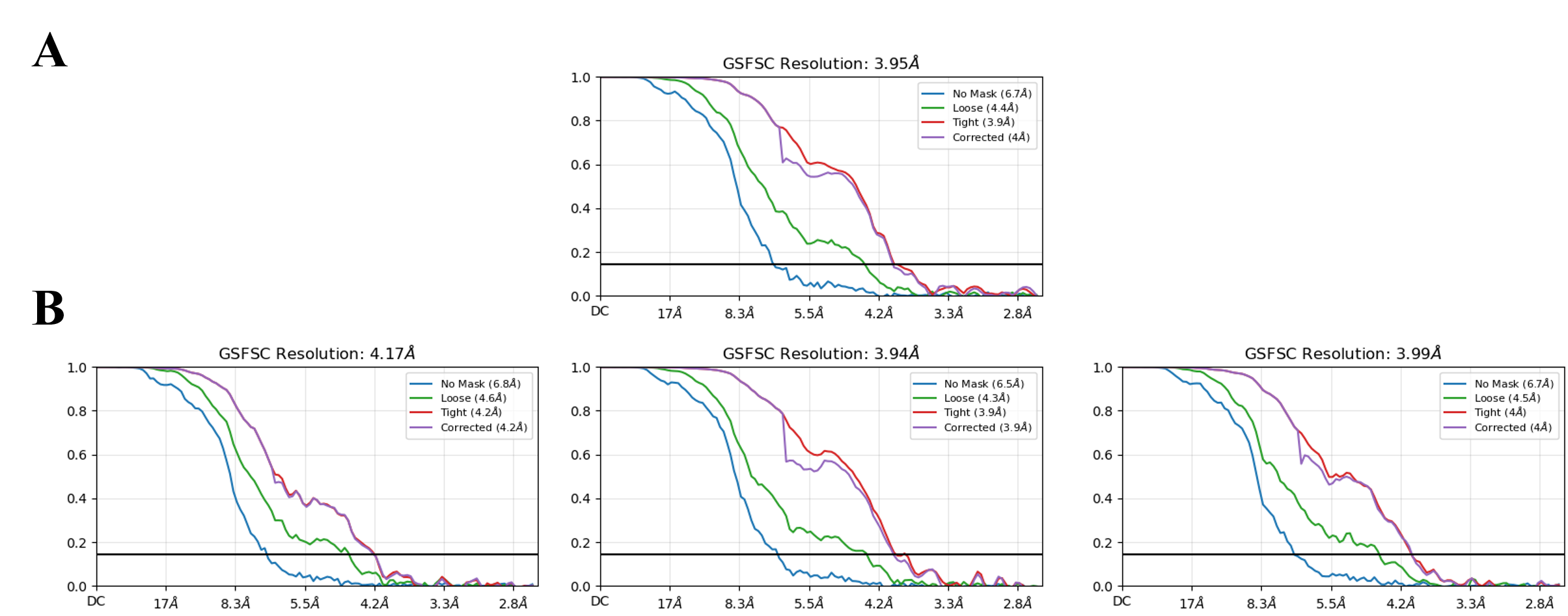}
    \caption{FSC curves obtained from 3D reconstruction using particles extracted with different methods on the EMPIAR-10081 dataset. \textbf{A}: The FSC curve for CryoPPP. \textbf{B}: From left to right, FSC curves obtained using different center-finding algorithms are presented: morphology and contour finding, the Crocker-Grier algorithm, and Non-Maximum Suppression.}
    \label{fig:10081_fsc}
\end{figure}

From the table, we observe that performance on the EMPIAR-10081 dataset is slightly better than that on the EMPIAR-10017 dataset. This difference is likely due to the fact that the EMPIAR-10017 dataset contains more densely connected and overlapping particles, and the particle distribution varies significantly across different regions, making center finding more challenging (as illustrated in \Cref{fig:nms}). Nonetheless, all center finding algorithms perform well, with precision, recall, and F1 scores all exceeding 0.9 on both datasets. In addition, Non-Maximum Suppression and the Crocker-Grier algorithm achieve the best performance on EMPIAR-10017 and EMPIAR-10081, respectively, and are therefore chosen as our default center finding methods in the main text. We further extract the particles identified by these algorithms and use them for 3D reconstruction; the corresponding results are shown in \Cref{fig:post_fsc_fig} and \Cref{fig:10081_fsc}. These results confirm that Non-Maximum Suppression and the Crocker-Grier algorithm indeed yield the highest 3D resolution, which is consistent with the mAP metric and justifies the use of mAP as a proxy for selecting the hyperparameters.


\begin{table}[hbt!]
	\centering
	\caption{Performance metrics for DeeLabV3 on the testing set of the synthetic EMPIAR-10017 dataset.}
	\label{table:simulation2}
\begin{tabular}{|c|c|c|c|c|}
\hline
\textbf{Architecture}                                            & \textbf{IoU}   & \textbf{Recall} & \textbf{F1 Score} & \textbf{Parameters} \\ \hline
DeepLabV3  &  0.9201          &  0.9560           &  0.9584               & 39.6M           \\ \hline
\end{tabular}
\end{table}

\begin{figure}[h]
    \centerline{\includegraphics[width=0.8\columnwidth]{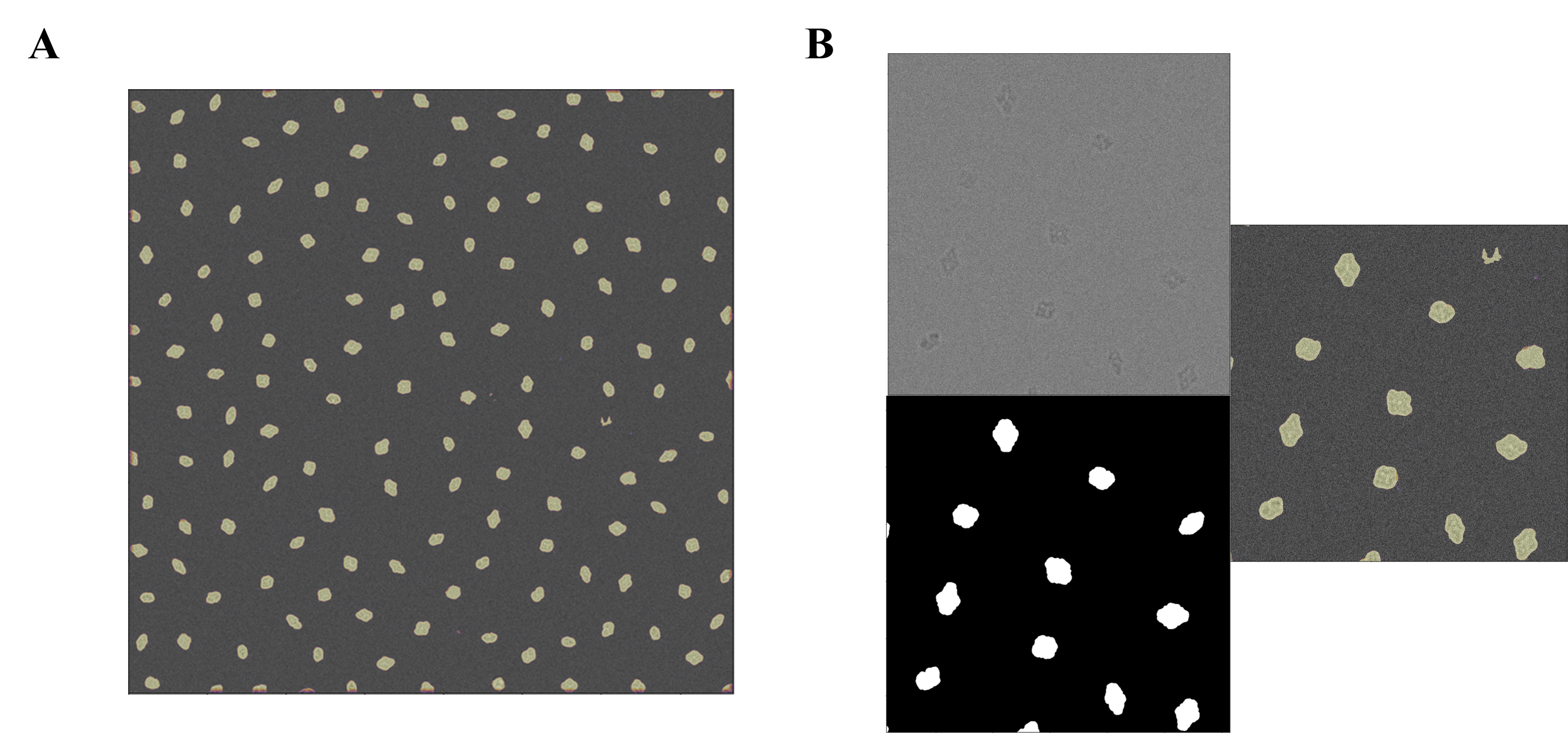}}
	\caption{Results of the synthetic EMPIAR-10017 experiment. \textbf{A}: Predicted segmentation map overlaid on the original micrograph from \textit{CRISP} using DeepLabV3. \textbf{B}: The left panel displays a noisy micrograph patch alongside its corresponding ground truth segmentation map, while the right panel shows a zoomed-in view of the predicted map from \textbf{A}.}
	\label{fig:sim_deeplab}
\end{figure}

\begin{figure}[h]
    \centerline{\includegraphics[width=0.8\columnwidth]{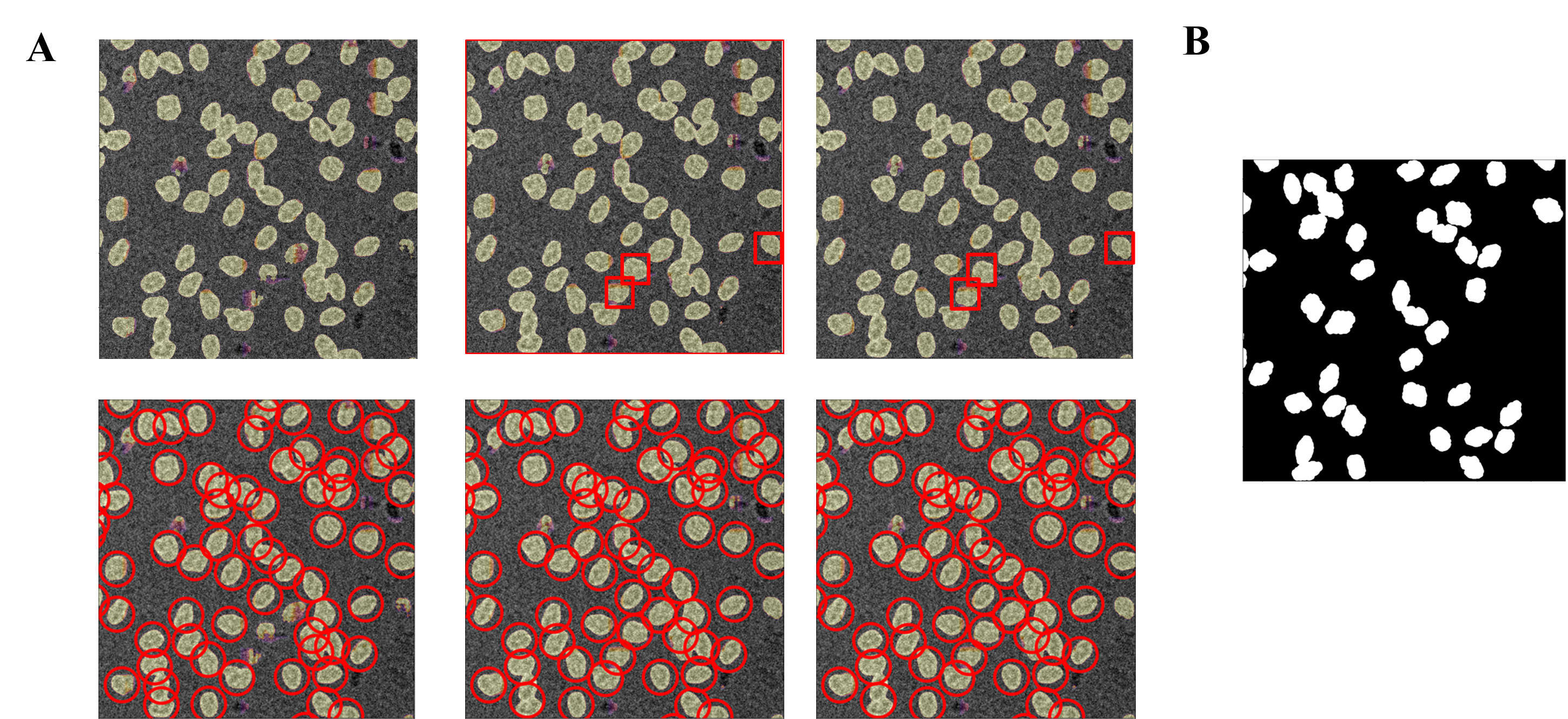}}
	\caption{Results of the EMPIAR-10017 experiment using DeepLabV3. \textbf{A}: From left to right, the predicted segmentation maps overlaid on the original micrograph are shown for the original model, the model with CRF, and the model with CD-CRF, respectively. The first row presents the raw prediction maps with a red square box highlighting the main differences, while the second row shows the corresponding maps after applying the center-finding algorithms, with a red circle indicating the selected particles. The full prediction maps are available in \Cref{fig:10017_deeplab_full}. \textbf{B}: The corresponding ground truth segmentation map.}
	\label{fig:10017_deep_lab}
\end{figure}

\subsection{Experiments with different segmentaion architectures}
In this section, we demonstrate the flexibility of our framework in switching between segmentation models and examine the robustness of our pipeline. We evaluate performance using both the synthetic and real EMPIAR-10017 datasets, as done previously with U-Net++. For these experiments, we use DeepLabV3 — a segmentation model from a different category that features a powerful encoder and employs simple interpolation in the output. DeepLabV3 is widely used in recent years \cite{s_le2021regularized} and has also been applied in cryo-EM studies \cite{s_zhang2019pixer}. All training parameters and the loss function remain identical to those used with U-Net++.

\begin{figure}[h]
    \centerline{\includegraphics[width=0.8\columnwidth]{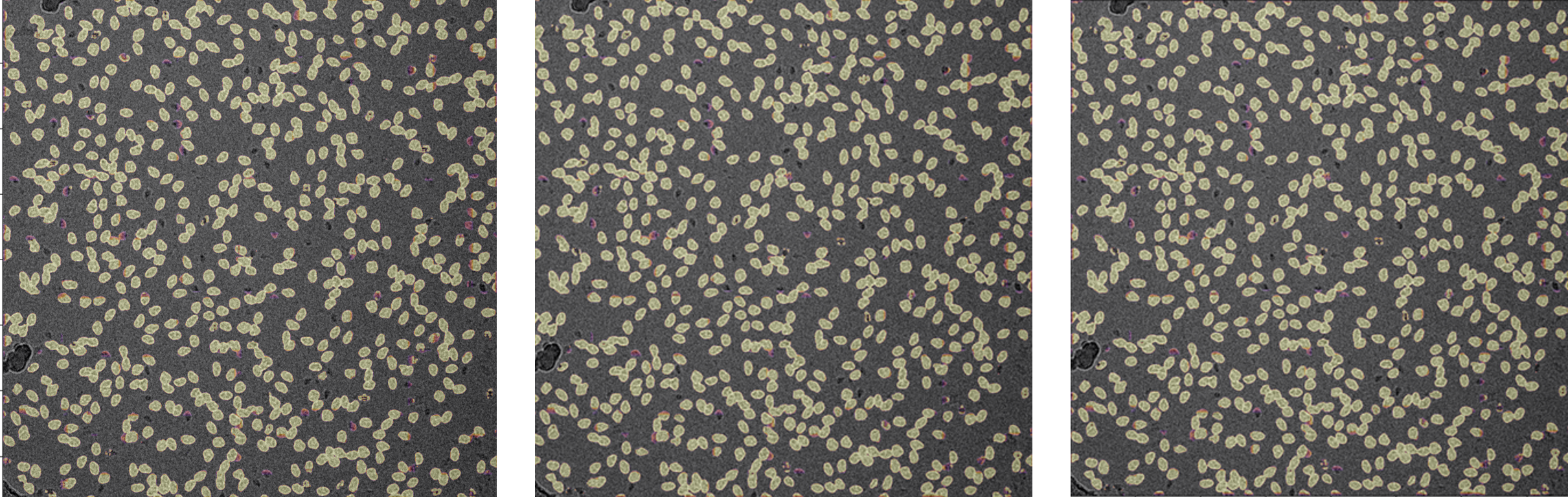}}
	\caption{Example of a full predicted segmentation map overlaid on the original micrograph from the real EMPIAR-10017 dataset. From left to right, the panels display the predicted segmentation maps obtained using the original DeepLabV3 model, the DeepLabV3 model with CRF, and the DeepLabV3 model with CD-CRF.}
	\label{fig:10017_deeplab_full}
\end{figure}

We first apply DeepLabV3 on the synthetic EMPIAR-10017 dataset; the results are summarized in \Cref{table:simulation2} and illustrated in \Cref{fig:sim_deeplab}. Although DeepLabV3's performance is slightly lower than that of U-Net++, all metrics — including IoU, Recall, and F1 score — exceed 0.9, and the predicted segmentation maps closely resemble the ground truth. Next, we apply the DeepLabV3 architecture to the real EMPIAR-10017 dataset. The corresponding results are shown in \Cref{fig:10017_deep_lab}, \Cref{table:metrics_crf_solver}, and \Cref{fig:10017_deeplab_full}. For this real dataset, we evaluate three configurations: the original model, the model with CRF, and the model with CD-CRF. As seen in \Cref{fig:10017_deep_lab} and \Cref{table:metrics_crf_solver}, both the CRF and CD-CRF configurations restore some broken particles and improve the metrics. 

\begin{figure}[h]
    \centering
    \includegraphics[width=0.95\columnwidth]{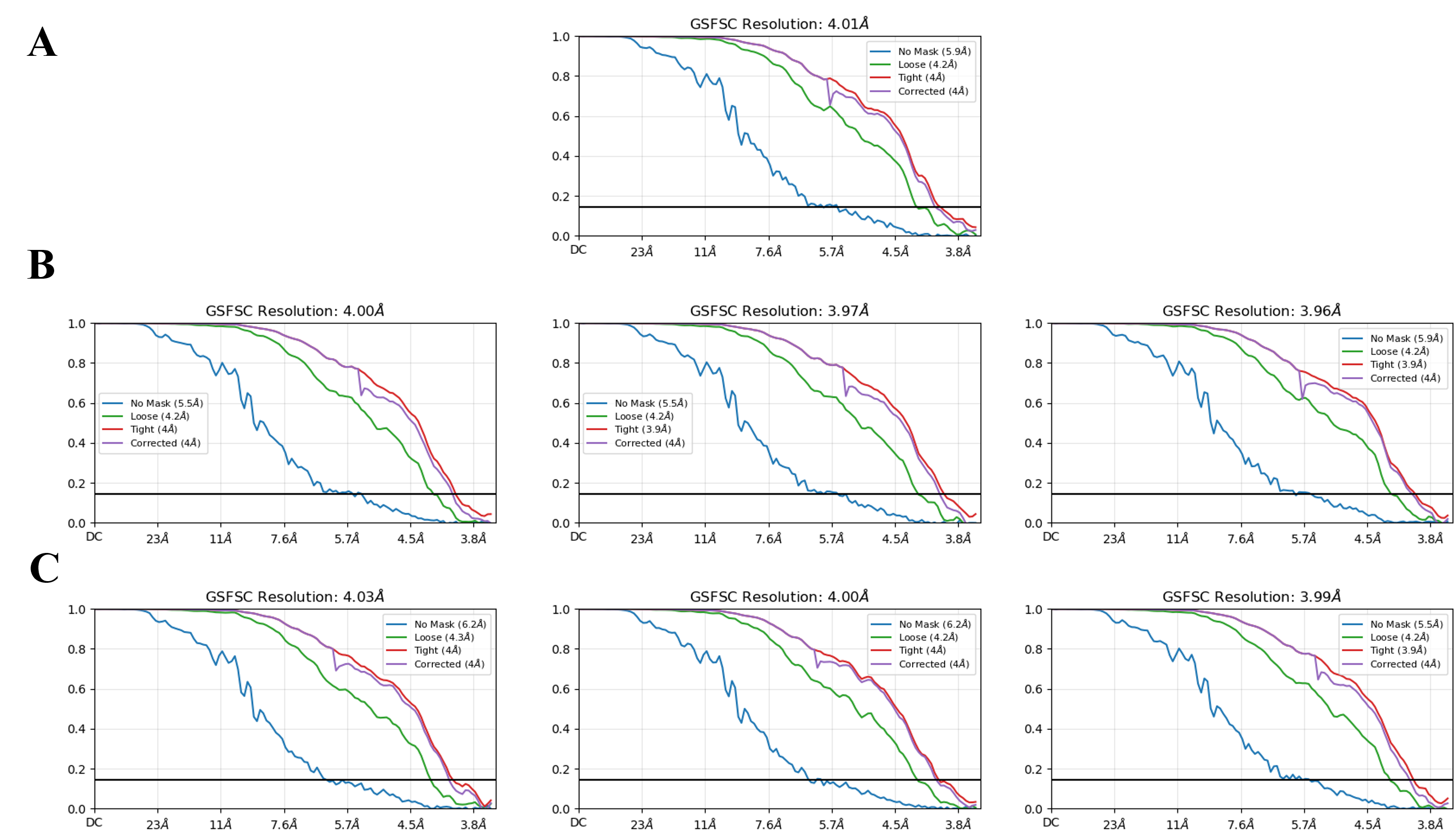}
    \caption{FSC curves obtained from 3D reconstruction using particles extracted with different methods on the EMPIAR-10017 dataset. \textbf{A}: The FSC curve for CryoPPP. \textbf{B}: From left to right, the FSC curves for U-Net++, U-Net++ with CRF, and U-Net++ with CD-CRF. \textbf{C}: From left to right, the FSC curves for DeepLabV3, DeepLabV3 with CRF, and DeepLabV3 with CD-CRF.}
    \label{fig:fsc_10017_unet}
\end{figure}

Moreover, we conduct downstream analysis for particle picking using our center finding algorithm with the hyperparameter selection procedure. The particles extracted by these methods are used for 3D reconstruction, and the results are presented in \Cref{fig:fsc_10017_unet}C. When CRF or CD-CRF is incorporated into the segmentation model, the final 3D resolution is comparable to that obtained with human-curated particles in CryoPPP. Overall, these experiments demonstrate that, with high-quality training datasets, the incorporation of CRF and the proposed postprocessing methods enables both popular architectures (U-Net++ and DeepLabV3) to achieve state-of-the-art performance.

\section{Comparsion with othe approaches}\label{sup:comparing}
In this section, we compare \textit{CRISP} with other methods in the particle picking task, where the performance is measured by the final FSC resolution value after 3D reconstruction. To facilitate a quick comparison with other state-of-the-art approaches trained on the CryoPPP dataset, we have collected and organized the results from \cite{s_gyawali2024cryosegnet,s_zhang2025upicker}\footnote{Since most state-of-the-art methods have relatively larger model sizes (often by an order of magnitude or more than ours) and differing training styles — some require training on a large variety of datasets while others necessitate self-supervised pretraining followed by fine-tuning — the computational resources required are formidable. It is also challenging to fairly compare these methods due to these differences; therefore, we also include the performance of the 3D reconstruction of the picked particles from the curated CryoPPP dataset as a reference benchmark.}. Given that training styles and model sizes differ among these methods, we also report the performance of 3D reconstructions from particles from CryoPPP, which, as a curated dataset generated by experts, serves as a high-quality reference.

The methods we compare include the widely used object detection–based pickers, such as CrYOLO, Topaz, and CryoTransformer\footnote{Upicker does not provide results on EMPIAR-10017 and is therefore not included in the comparison.}, as well as state-of-the-art image segmentation–based particle pickers including CASSPER and CryoSegNet. The resolution and the number of particles picked by each algorithm on the EMPIAR-10017 and EMPIAR-10081 datasets are presented in \Cref{table:full_compare}. From the table, we observe that \textit{CRISP} picks a reasonable number of particles compared with other approaches, and its resolution outperforms all the other pickers. Moreover, the final 3D resolution from the particles picked by \textit{CRISP} is comparable to that achieved using the curated CryoPPP dataset. 

It should be noted that the CryoPPP particles were selected by experts and underwent several rigorous cleaning processes. In contrast, \textit{CRISP} does not require any cleaning (such as 2D or 3D classification) because the 3D volume is directly reconstructed from the picked particles, indicating that the particles identified by \textit{CRISP} are already of high quality and can significantly reduce the time needed for data cleaning. Finally, it should be emphasized that \textit{CRISP} is not limited to particle picking; the predicted segmentation map from our segmentation pipeline can potentially serve other downstream tasks, and the framework itself allows users to rapidly explore different components and quickly evaluate them using various metrics.

\begin{table*}[hbt!]
	\centering
	\caption{Comparison of the number of particles and corresponding 3D reconstruction resolutions across different state-of-the-art methods on the EMPIAR-10017 and EMPIAR-10081 datasets. The compared algorithms include both object-detection–based and image segmentation–based pickers (details in \Cref{sup:back}). The expert-curated dataset is also included for reference.}
	\label{table:full_compare}
    {\small
	\begin{tabular}{|c|c|c|c|c|c|c|c|}
 	\hline
		\textbf{EMPIAR-10017} & \textbf{CrYOLO} & \textbf{Topaz} & \textbf{CryoTransformer} & \textbf{CASSPER} & \textbf{CryoSegNet} & \textbf{\textit{CRISP}} & \textbf{CryoPPP} \\
		\hline
		\textbf{Number of particles} & 47,704 & 45,511 & 43,735 &  38,460 & 10,026 & 49,517 & 49,391 \\
		\textbf{Resolution} & 4.87 &  5.09 & 5.61 & 5.33 & 6.91 & \textbf{3.94}  & 3.95 \\
        \hline
		\textbf{EMPIAR-10081} & \textbf{CrYOLO} & \textbf{Topaz} & \textbf{CryoTransformer} & \textbf{CASSPER} & \textbf{CryoSegNet} & \textbf{\textit{CRISP}} & \textbf{CryoPPP}\\
		\hline
		\textbf{Number of particles} & 36,821 & 37,808 & 88,632 & 27,299 & 44,891 & 42,657  & 39,352 \\
		\textbf{Resolution} & 5.38 &  5.10 & 5.47 & 5.79 & 4.18 & \textbf{3.96}  & 4.01 \\
		\hline
	\end{tabular}
   }
\end{table*}



\section{Data preparation} \label{sup:setup}

\subsection{Data preparation of the synthetic EMPIAR-10017 dataset}
\begin{figure}[h]
    \centerline{\includegraphics[width=0.8\columnwidth]{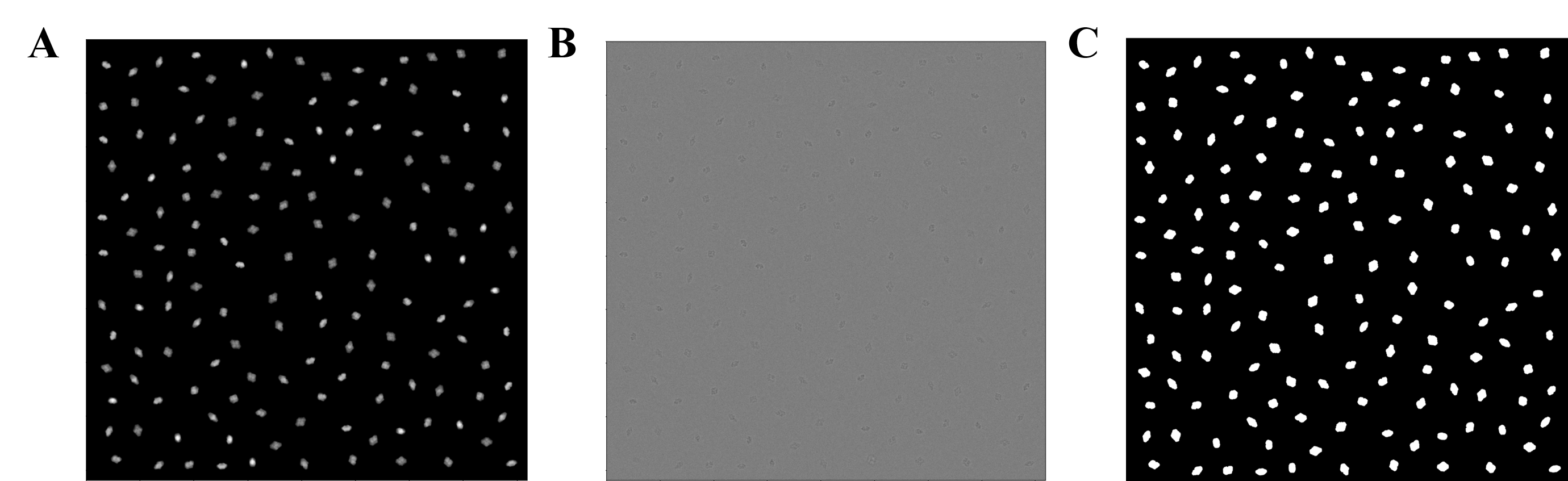}}
    \caption{Example micrograph from the synthetic EMPIAR-10017 dataset generated by our framework. \textbf{A}: The clean synthetic micrograph. \textbf{B}: The corresponding noisy synthetic micrograph. \textbf{C}: The corresponding ground truth segmentation map. }
    \label{fig:simu_ex_fig}
\end{figure}

We generate a simulated dataset according to our synthetic label generation pipeline described in the main text. First, we download the 3D density map and experimental information — including the Contrast Transfer Function (CTF) parameters — from EMPIAR-10017. This dataset, which features Beta-galactosidase collected using Falcon-II, serves as the basis for our simulation. The defocus value is randomly sampled from the metadata in the real experiment, and the orientation is randomly sampled from \(SO(3)\). Based on these parameters, we use \code{ASPIRE} \cite{s_Aspire} to project the 3D volume into particle images. These images are then multiplied with CTF in the frequency domain and corrupted with Gaussian noise. A total of 84 micrographs, each of size \(4{,}096 \times 4{,}096\) and containing approximately 150 2D projections per micrograph, are generated. The SNR is adjusted to 0.005 to reflect the high noise levels observed in real micrographs. Representative examples of a clean micrograph and a noisy micrograph can be found in \Cref{fig:simu_ex_fig}A and B, respectively.

To obtain the corresponding label, we apply Li's thresholding method \cite{s_li1998iterative} to each clean micrograph for binarization. A representative label segmentation map is shown in \Cref{fig:simu_ex_fig}C, where the shapes and positions of the 2D projections in the segmentation map clearly match those in the clean micrograph. To test the limits of our framework, we fix the training set and validation set to contain only 16 and 6 micrographs, respectively.

\subsection{Data preparation of the EMPIAR-10017 dataset}
We selected the EMPIAR-10017 dataset for testing because it is widely used in the particle picking literature and, with a storage requirement of only 0.005 TB, it is relatively small compared to other CryoPPP datasets. This makes it suitable for conducting various experiments on a mainstream GPU.

For this dataset, we first download the micrographs and the corresponding particle coordinates from CryoPPP. The dataset comprises 84 micrographs with a pixel size of 1.77 \AA{} and a resolution of \(4{,}096 \times 4{,}096\) pixels per micrograph. We extract particles from each micrograph and use these particles to reconstruct a 3D density map. After reconstruction, the orientation of each particle is obtained, and the 3D mask of the density map is used to generate reprojected images according to these orientations using RELION \cite{s_2D_clf_relion}. 

\begin{figure}[h]
    \centerline{\includegraphics[width=0.8\columnwidth]{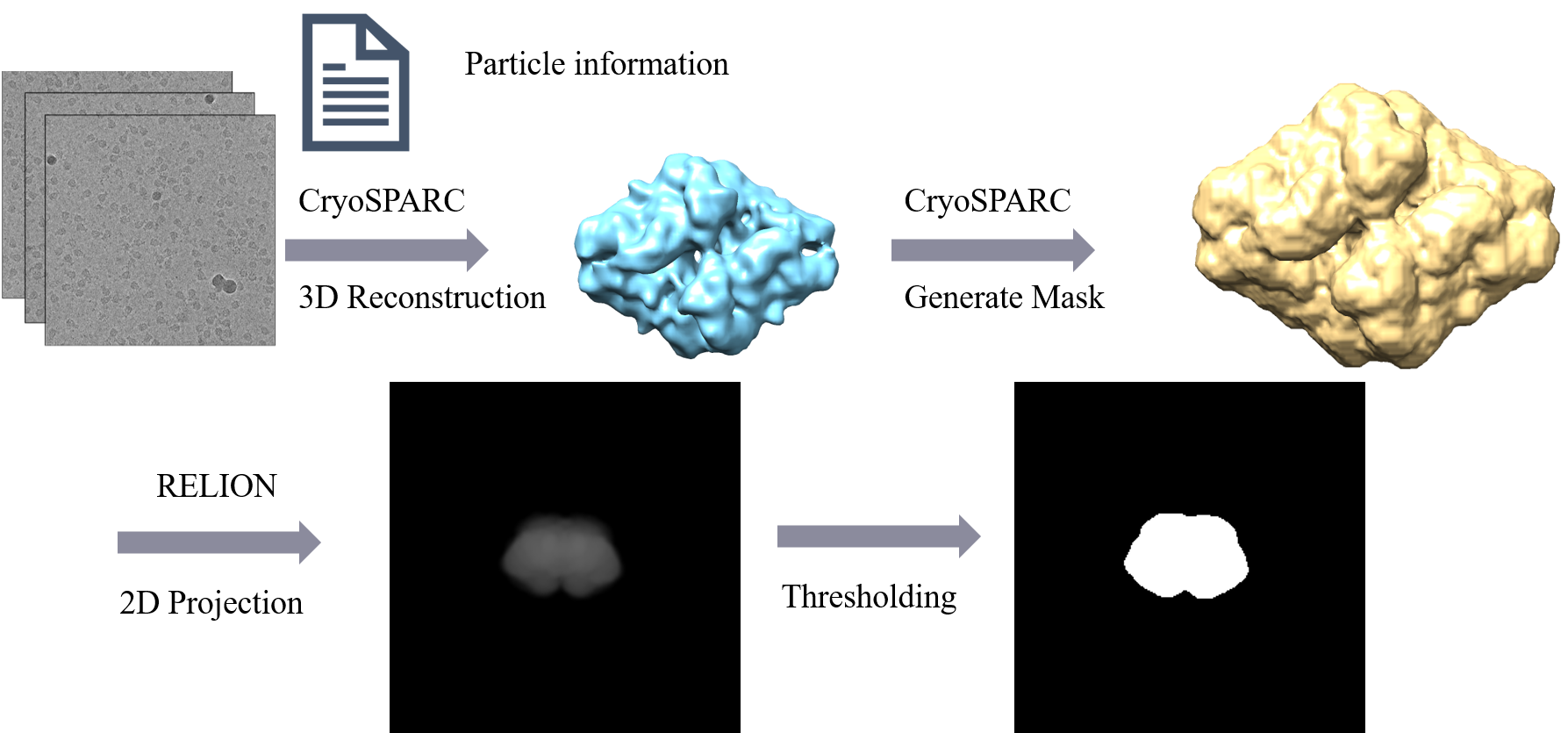}}
	\caption{Flowchart of the workflow for generating the ground truth segmentation map for a real dataset.}
	\label{fig:synthetic_flow}
\end{figure}

\begin{figure}[h]
    \centering
    \includegraphics[width=0.8\columnwidth]{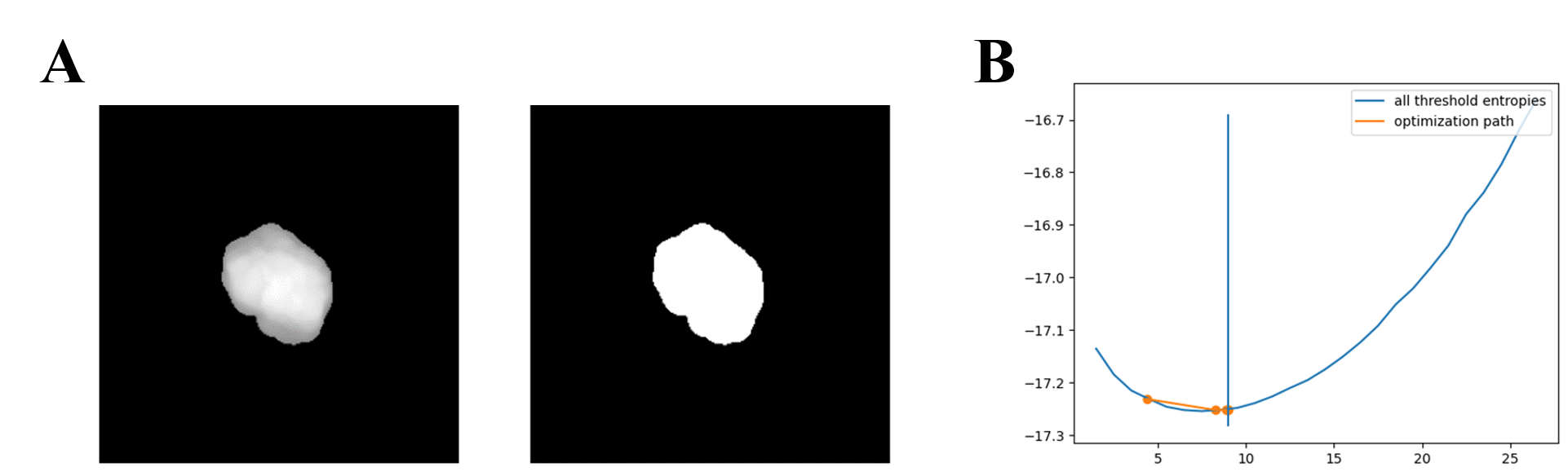}
    \caption{\textbf{A}: The left figure shows a typical reprojection image using the 3D mask obtained with \Cref{fig:synthetic_flow}, while the right figure illustrates the corresponding thresholding results. \textbf{B}: Visualization of the optimization process of Li's thresholding algorithm, where the $x$-axis represents the threshold values and the $y$-axis corresponds to the cross-entropy computed between the foreground pixels and their mean, as well as between the background pixels and their mean.}
    \label{fig:thresholding_fig}
\end{figure}

\begin{figure}[h]
    \centering
    \includegraphics[width=0.5\columnwidth]{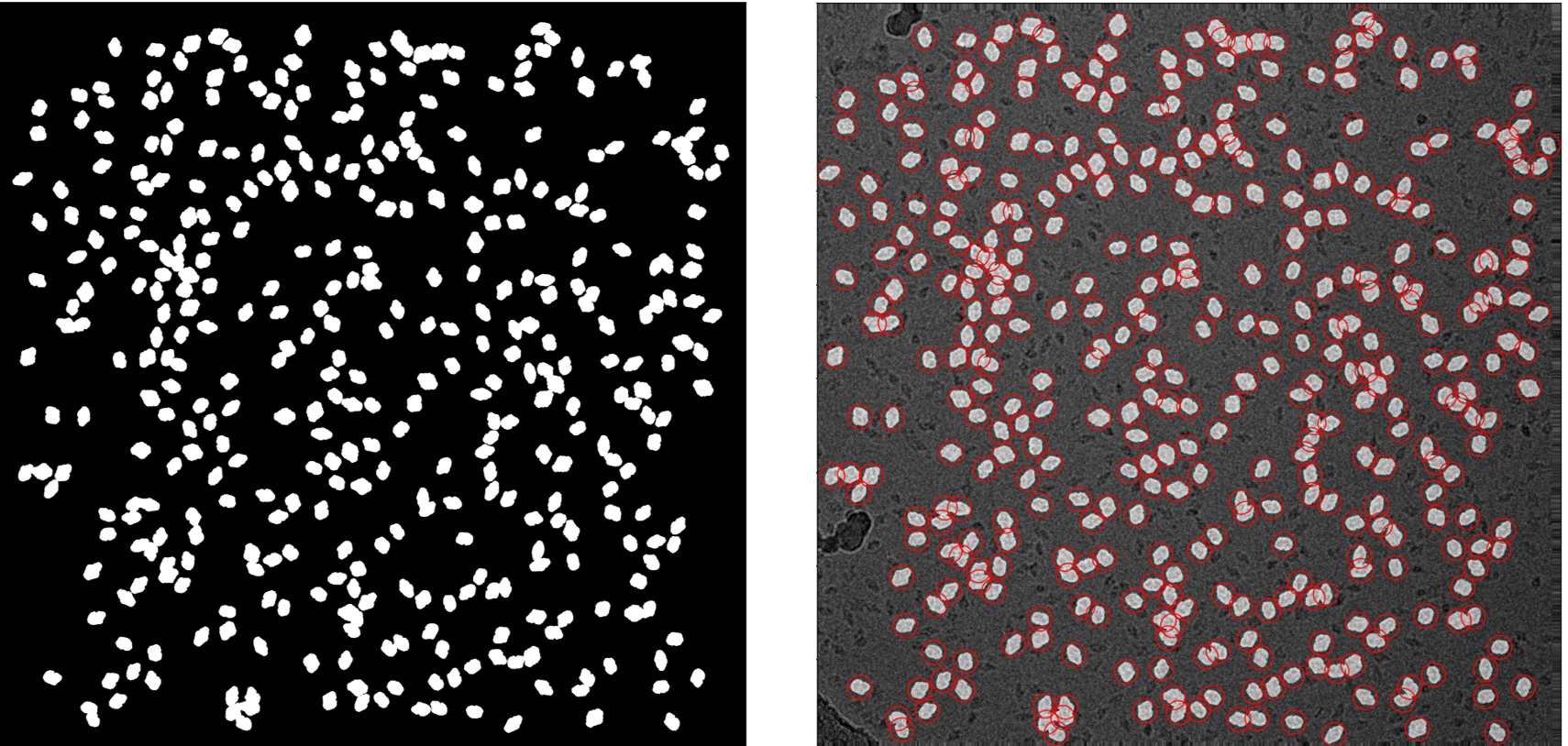}
    \caption{Example ground truth segmentation map from the EMPIAR-10017 dataset generated by our framework. \textbf{A}: The ground truth segmentation map obtained by stitching reprojection images together. \textbf{B}: An overlay of \textbf{A} on the original micrograph, where the red box indicates the particle locations provided by CryoPPP.}
    \label{fig:real_ex_fig}
\end{figure}

Subsequently, Li's thresholding method\footnote{In \textit{CRISP}, users are free to choose from various popular thresholding methods, including both global and local thresholding techniques available in the \code{scikit-image} library \cite{s_van2014scikit}. Empirically, we found that Li's thresholding works best.} is employed to obtain binary segmented particles. The optimization process for thresholding is illustrated in \Cref{fig:thresholding_fig}, and the entire process is summarized in \Cref{fig:synthetic_flow}. Finally, we generate the segmentation map by placing each binary segmented particle into the micrograph according to the original particle coordinates. The resulting segmentation map, shown in \Cref{fig:real_ex_fig}, closely matches the ground truth bounding boxes. We then perform normalization using \code{Topaz} \cite{s_bepler2019positive} and fix the training and validation sets to contain only 16 and 6 micrographs, respectively, with the remaining 62 micrographs designated as the test set.

\subsection{Data preparation of the EMPIAR-10081 dataset}
For the real EMPIAR-10081 dataset, we first download the micrographs and the corresponding particle coordinates from CryoPPP \cite{s_dhakal2023large}. This dataset comprises 300 micrographs with a pixel size of 1.3 \AA, featuring a membrane protein, and each micrograph has dimensions of \(3{,}710 \times 3{,}838\). It is selected because it poses a greater challenge for particle picking. Membrane proteins must be extracted from their native lipid bilayers —typically using detergents or by reconstituting them into lipid nanodiscs —which introduces a surrounding detergent micelle or lipid environment. This additional density can obscure the protein’s features and lower the overall contrast between the protein and the background, resulting in a reduced SNR ratio that complicates particle picking. Notably, among the CryoPPP datasets, EMPIAR-10081 is chosen because it is relatively small and thus computationally manageable.

We follow a procedure similar to that used for the EMPIAR-10017 datasets. First, particles are extracted from each micrograph and used to reconstruct a 3D density map. After reconstruction, the orientation of each particle is determined, and the 3D mask of the density map is used to generate reprojected images according to these orientations using RELION. Next, Li's thresholding method is employed to obtain binary segmented particles. Finally, the segmentation map is generated by placing each binary segmented particle into the micrograph according to the original particle coordinates. We then perform normalization using \code{Topaz} and fix the training and validation sets to contain only 16 and 6 micrographs, respectively, with the remaining 278 micrographs designated as the test set.

\subsection{Training details} \label{sup:analyze}
After splitting the data into training, validation, and test sets, we perform center cropping on the EMPIAR-10017 dataset to obtain images of size \(3{,}840 \times 3{,}840\) pixels, since the boundaries of some micrographs contain artifacts. During training, four patches of size \(512 \times 512\) are extracted from each micrograph within a batch and fed into the segmentation model. The model outputs patches of the same size, which are then compared with the corresponding regions in the ground truth segmentation mask to compute performance metrics.

\begin{figure}[h]
    \centerline{\includegraphics[width=0.4\columnwidth]{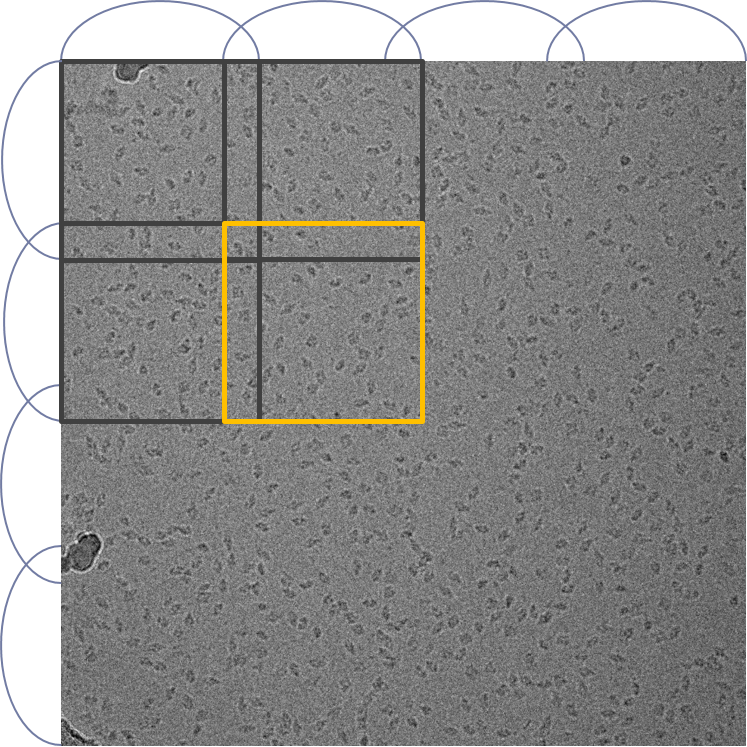}}
    \caption{Illustration of the patch division scheme used during validation and testing. Specifically, the micrographs are divided into overlapping patches using a sliding window of size $512 \times 512$, spanning from the top left to the bottom right corner.}
	\label{fig:patch}
\end{figure}

In the validation and testing stages, each micrograph is divided into overlapping patches of size \(512 \times 512\) with an overlap of 64 pixels between adjacent patches. The output patches are subsequently stitched together to reconstruct the full segmentation map, as depicted in \Cref{fig:patch}. Specifically, each image patch is multiplied element‐wise by a weight map\footnote{The weight map is a Gaussian-like weight map constructed by first creating a one-dimensional ramp that smoothly increases from zero to one over a specified bandwidth, concatenating it with a flat region and a mirrored decrease, and then forming a two-dimensional map via the outer product of this ramp with itself. This produces a spatial weighting that emphasizes the center of each patch while tapering off smoothly toward the edges.} before being placed into the overall image grid, and overlapping regions are combined using a weighted average. This weighting scheme effectively blends the patches to mitigate boundary artifacts and yield a seamless, high-quality composite image.

The batch size is set to 2 to accommodate a mainstream GPU, and the training is conducted for 50 epochs. We use the Adam optimizer with a learning rate of 0.001 for all experiments. Additionally, One-Cycle Learning is employed for learning rate scheduling to facilitate super-convergence and mitigate the risk of overfitting \cite{s_smith2019super}. When fine-tuning the CRF with the CNN, we follow the strategy described in \cite{s_le2021regularized} by employing a discriminative learning rate scheme, where the CNN is trained with a learning rate of \(10^{-5}\) and the CRF with a learning rate of \(10^{-3}\) for 50 epochs. The training parameters are summarized in \Cref{table:training_hyper}.

\begin{table}[!hbt]
	\centering
	\caption{Training hyperparameters used in the segmentation pipeline across all experiments in this study.}
	\label{table:training_hyper}
\begin{tabular}{|c|c|}
\hline
                         & Hyperparameters               \\ \hline
Patch   Size              & 512 $\times$ 512 \\ \hline
Batch   Size             & 2   mic (4   patches)         \\ \hline
Epoch                    & 50                            \\ \hline
Optimizer                & Adam                          \\ \hline
Learning   rate          & 0.001                         \\ \hline
Learning   rate scheduler & One-Cycle Learning                  \\ \hline
\end{tabular}
\end{table}

\subsection{Evaluation metrics} \label{sup:metrics}

Our segmentation model assigns a class label to each pixel in an image, and its performance is evaluated by comparing the predicted segmentation with the ground truth mask on a pixel-wise basis. In this context, every pixel is classified as either belonging to a target (positive, particle) class or not (negative, background). The quality of the segmentation is quantified using metrics derived from the confusion matrix, which is constructed from the following four fundamental quantities:
\begin{itemize}
    \item \textbf{True Positives (TP):} the number of pixels correctly predicted as belonging to the target class.
    \item \textbf{False Positives (FP):} the number of pixels incorrectly predicted as belonging to the target class.
    \item \textbf{False Negatives (FN):} the number of pixels that belong to the target class but are misclassified as not belonging to it.
    \item \textbf{True Negatives (TN):} the number of pixels correctly predicted as not belonging to the target class.
\end{itemize}

Based on these definitions, the following metrics are used in our evaluation. The IoU, also known as the Jaccard Index, measures the spatial overlap between the predicted segmentation and the ground truth mask. It is defined as
\[
\text{IoU} = \frac{|P \cap G|}{|P \cup G|} = \frac{\text{TP}}{\text{TP} + \text{FP} + \text{FN}},
\]
where \(P\) and \(G\) denote the sets of pixels in the predicted and ground truth masks, respectively.

The \emph{Recall} (or sensitivity) quantifies the ability of the segmentation algorithm to detect all relevant pixels of the target class. It is given by
\[
\text{Recall} = \frac{\text{TP}}{\text{TP} + \text{FN}}.
\]
A high recall indicates that most target pixels are correctly detected.

The \emph{Precision} (or positive predictive value) measures the accuracy of the pixels predicted as belonging to the target class, and is defined as
\[
\text{Precision} = \frac{\text{TP}}{\text{TP} + \text{FP}}.
\]
A high precision reflects a low incidence of false detections.

The \emph{F1 Score} combines precision and recall into a single metric by taking their harmonic mean, balancing the trade-off between the two:
\[
\text{F1 Score} = 2 \times \frac{\text{Precision} \times \text{Recall}}{\text{Precision} + \text{Recall}} = \frac{2\,\text{TP}}{2\,\text{TP} + \text{FP} + \text{FN}}.
\]
The F1 Score is particularly useful when both false positives and false negatives are critical to the performance assessment. Each of these metrics is computed on a pixel-wise basis over the entire image, making them indispensable for quantitatively evaluating segmentation algorithms, particularly with respect to their sensitivity to over-segmentation (increased FP) and under-segmentation (increased FN).

For the 3D reconstruction, the Fourier Shell Correlation (FSC) curve is computed using the tight mask generated by cryoSPARC \cite{s_punjani2017cryosparc}. This method incorporates noise substitution as a correction technique, as detailed in \cite{s_chen2013high}. The procedure involves dividing particles into two random subsets to create two half-maps, enabling the calculation of the initial FSC. For each half-map, the phases beyond a certain resolution are randomized. After applying the tight mask to both half-maps, the FSC is computed by comparing the Fourier transforms of the two independent half-maps. In the FSC formula,
\[
\text{FSC}(s) = \frac{\sum_{i} F_1(s_i)\, F_2^*(s_i)}{\sqrt{\sum_{i} |F_1(s_i)|^2 \; \sum_{i} |F_2(s_i)|^2}},
\]
the index \(s_i\) denotes the individual Fourier coefficients (or voxels) lying within a thin spherical shell at spatial frequency \(s\); the summation is performed over all voxels whose spatial frequency magnitudes fall within the shell. The resulting FSC, in combination with the initial FSC calculated before phase randomization, facilitates the determination of the corrected FSC. This strategy, outlined in \cite{s_chen2013high}, compensates for the correlation effects induced by masking. The resolution of the map is determined at the point where the FSC drops below the 0.143 threshold \cite{s_rosenthal2003optimal}.

\subsection{3D reconstruction}
For the 3D reconstruction, we feed the particles selected by each method in \textit{CRISP} into cryoSPARC without any 2D cleaning. First, we perform ab-initio reconstruction, where an initial 3D model is created directly from the selected particles without using any prior structural model. Finally, homogeneous refinement is conducted using the ab-initio density map and the particles to enhance particle alignment accuracy and improve the overall resolution of the reconstructed 3D density map.

\subsection{Data availability}
The synthetic datasets were generated using custom code in our modular framework. In addition, the modular framework is available at the following GitHub repository: \href{https://github.com/phonchi/CryoParticleSegment/}{https://github.com/phonchi/CryoParticleSegment/}. The EMPIAR-10017 and EMPIAR-10081 datasets can be accessed via \href{https://github.com/BioinfoMachineLearning/cryoppp}{CryoPPP}.

\section{Runtime of the framework}
\label{chap:runtime}

The experiments were performed on a Google Colab instance featuring an Intel(R) Xeon(R) CPU @ 2.20GHz with 6 cores, 53 GB of RAM, and an NVIDIA L4 GPU. The training and CRF fine-tuning can be completed within 1 hour. For the particle picking process, the Crocker-Grier algorithm and non-maximum suppression require roughly 1 second per micrograph, while the morphology and contour-finding approach takes approximately 3 seconds per micrograph.


\clearpage


\section{Supplementary Figures}

\begin{figure}[h]
    \centerline{\includegraphics[width=0.8\columnwidth]{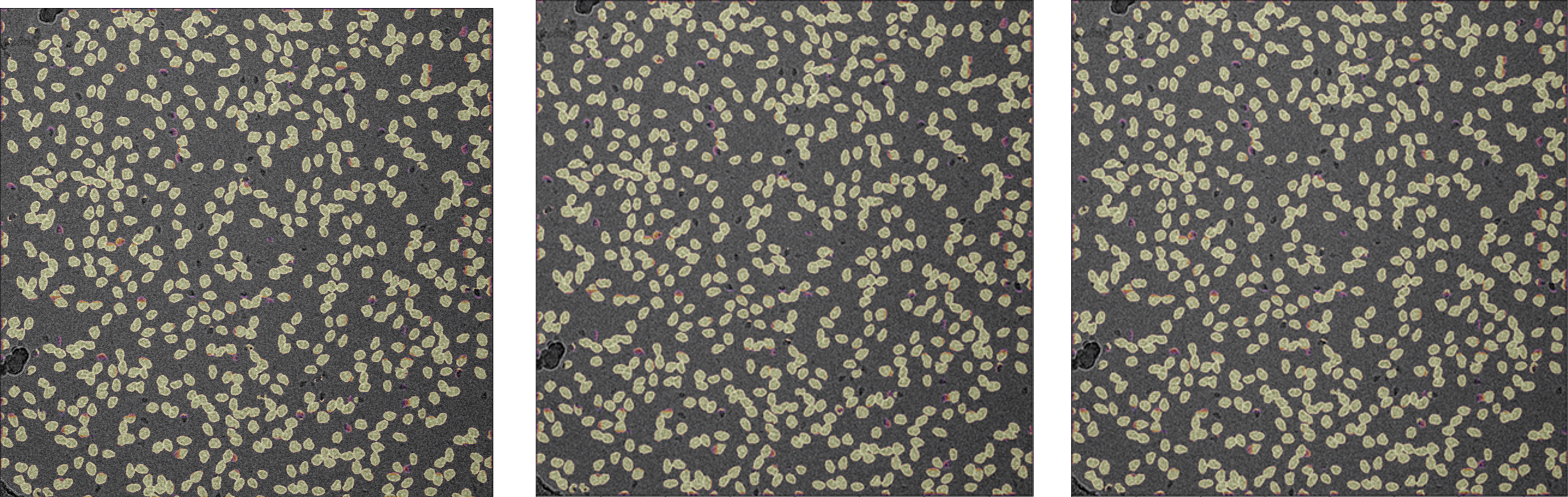}}
	\caption{Example of a full predicted segmentation map overlaid on the original micrograph from the real EMPIAR-10017 dataset. From left to right, the panels show the predicted segmentation maps obtained using the original U-Net++ model, the U-Net++ model with CRF, and the U-Net++ model with CD-CRF, respectively.}
	\label{fig:10017_full}
\end{figure}

\begin{figure}[h]
    \centering
    \includegraphics[width=0.8\columnwidth]{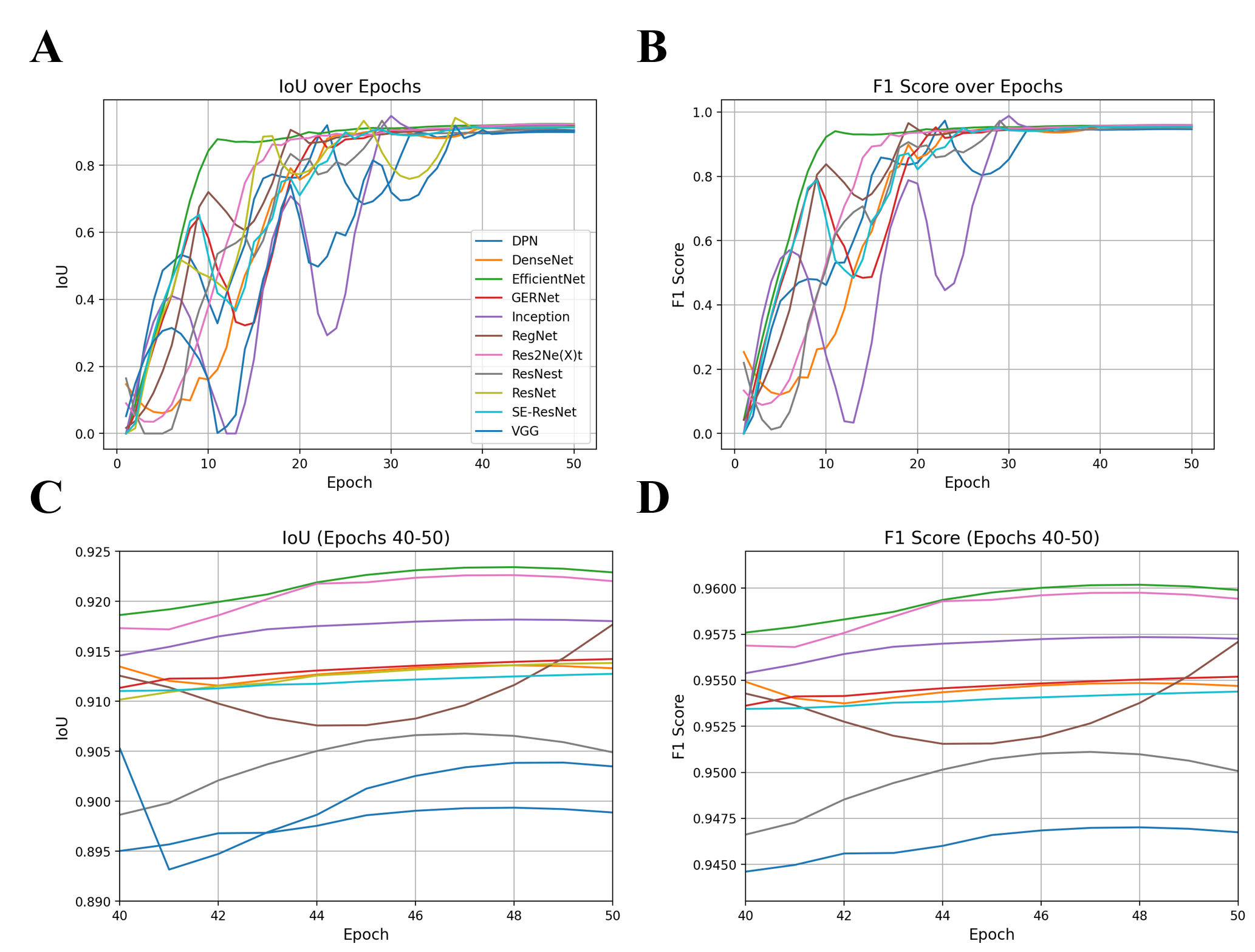}
    \caption{The training dynamics of various CNN models available in \textit{CRISP} as encoders on the synthetic EMPIAR-10017 dataset are presented. \textbf{A}: IoU versus epochs. \textbf{B}: F1 score versus epochs. \textbf{C}: A zoomed-in view of \textbf{A} from epochs 40 to 50. \textbf{D}: A zoomed-in view of \textbf{B} from epochs 40 to 50. Note that the Savitzky–Golay filter—a technique based on local polynomial regression—is applied to smooth the noisy performance metrics over successive epochs. Specifically, for each data point, the filter fits a second-order polynomial over a moving window of length 11, effectively reducing high-frequency noise while preserving the essential features of the underlying signal.}
    \label{fig:simu_metric2}
\end{figure}

\begin{figure}[h]
    \centering
    \includegraphics[width=0.55\columnwidth]{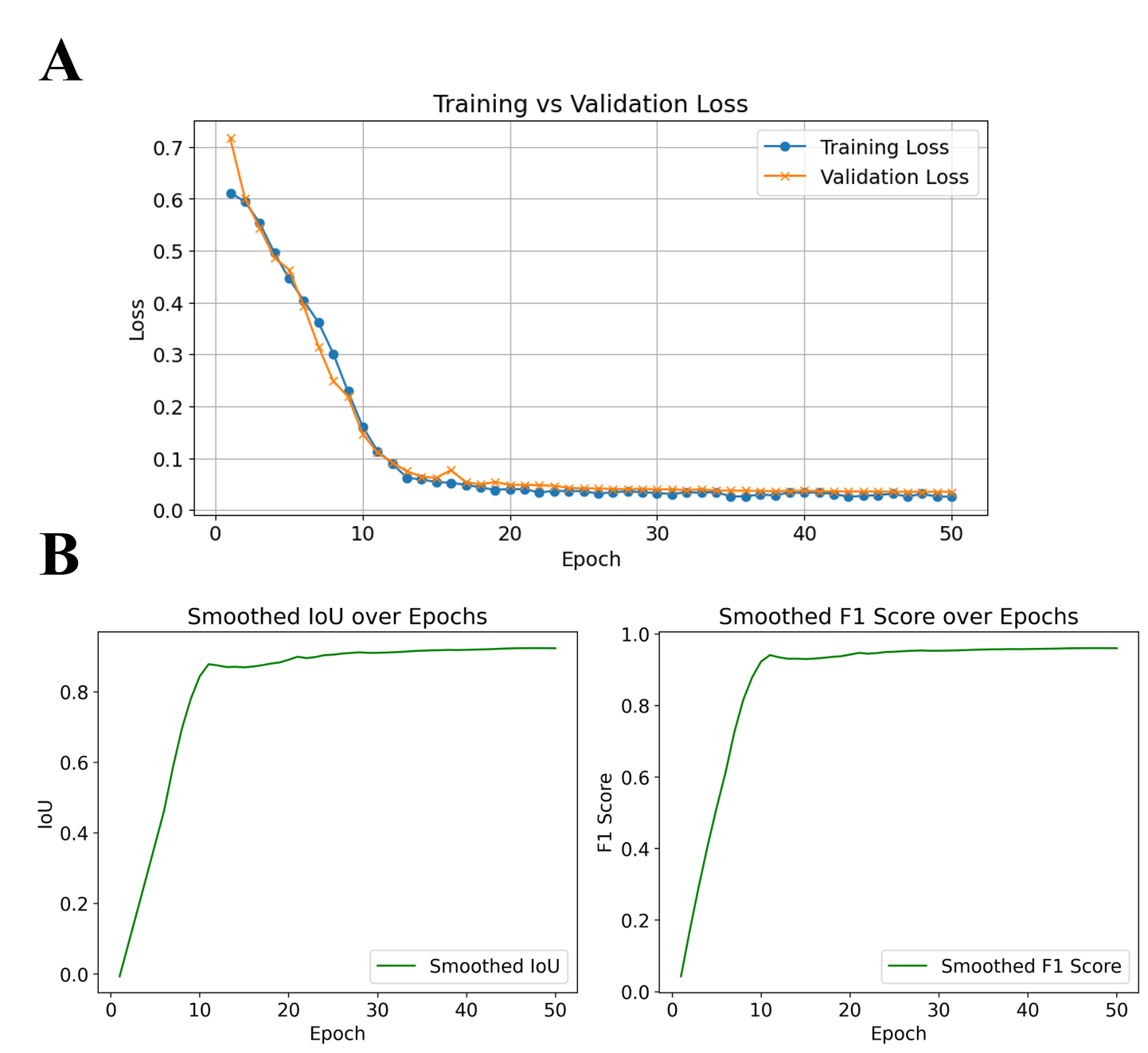}
    \caption{The training dynamics of \textit{CRISP} on the synthetic EMPIAR-10017 dataset are illustrated as follows: \textbf{A}: Training and validation loss versus epochs. \textbf{B}: IoU score versus epochs. \textbf{C}: F1 score versus epochs. Note that the Savitzky–Golay filter—a local polynomial regression method—is applied to smooth the noisy performance metrics across epochs by fitting a second-order polynomial over an 11-point moving window, effectively reducing high-frequency noise while preserving key signal features.}
    \label{fig:simu_metric}
\end{figure}

\begin{figure}[h]
    \centering
    \includegraphics[width=0.55\columnwidth]{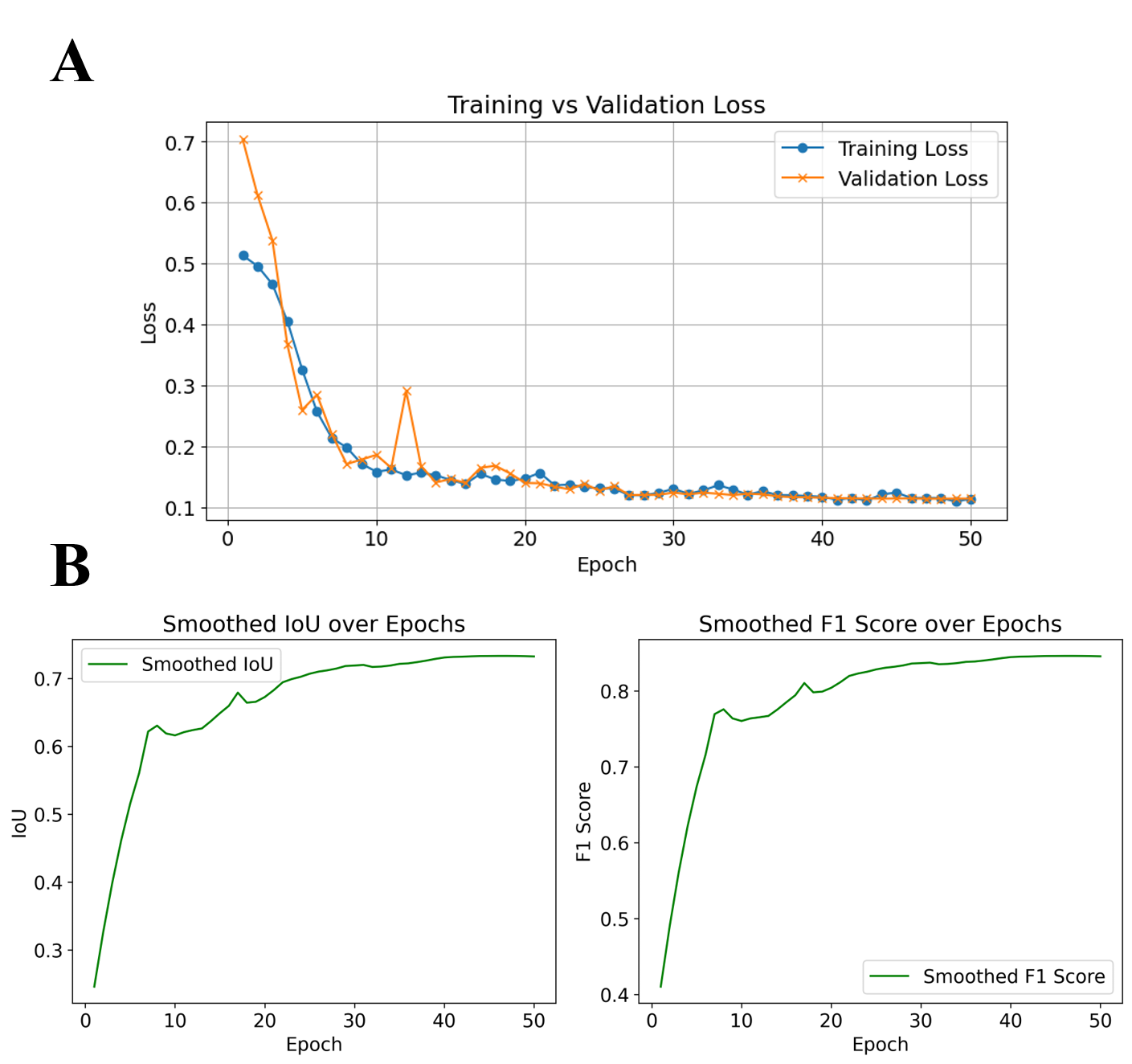}
    \caption{The training dynamics of \textit{CRISP} on the EMPIAR-10017 dataset are presented. \textbf{A}: Training and validation loss versus epochs. \textbf{B}: IoU score versus epochs. \textbf{C}: F1 score versus epochs. Note that the Savitzky–Golay filter—a local polynomial regression method—is applied to smooth the noisy performance metrics across epochs by fitting a second-order polynomial over an 11-point moving window, effectively reducing high-frequency noise while preserving key signal features.}
    \label{fig:10017_metric}
\end{figure}

\begin{figure}[h]
    \centerline{\includegraphics[width=0.3\columnwidth]{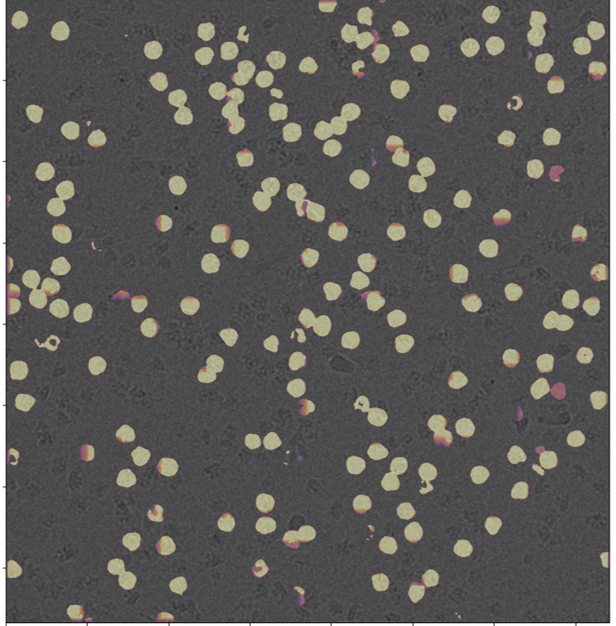}}
	\caption{Example of a full predicted segmentation map overlaid on the original micrograph from the real EMPIAR-10081 dataset.}
	\label{fig:10081_full}
\end{figure}

\begin{figure}[h]
    \centering
    \includegraphics[width=0.55\columnwidth]{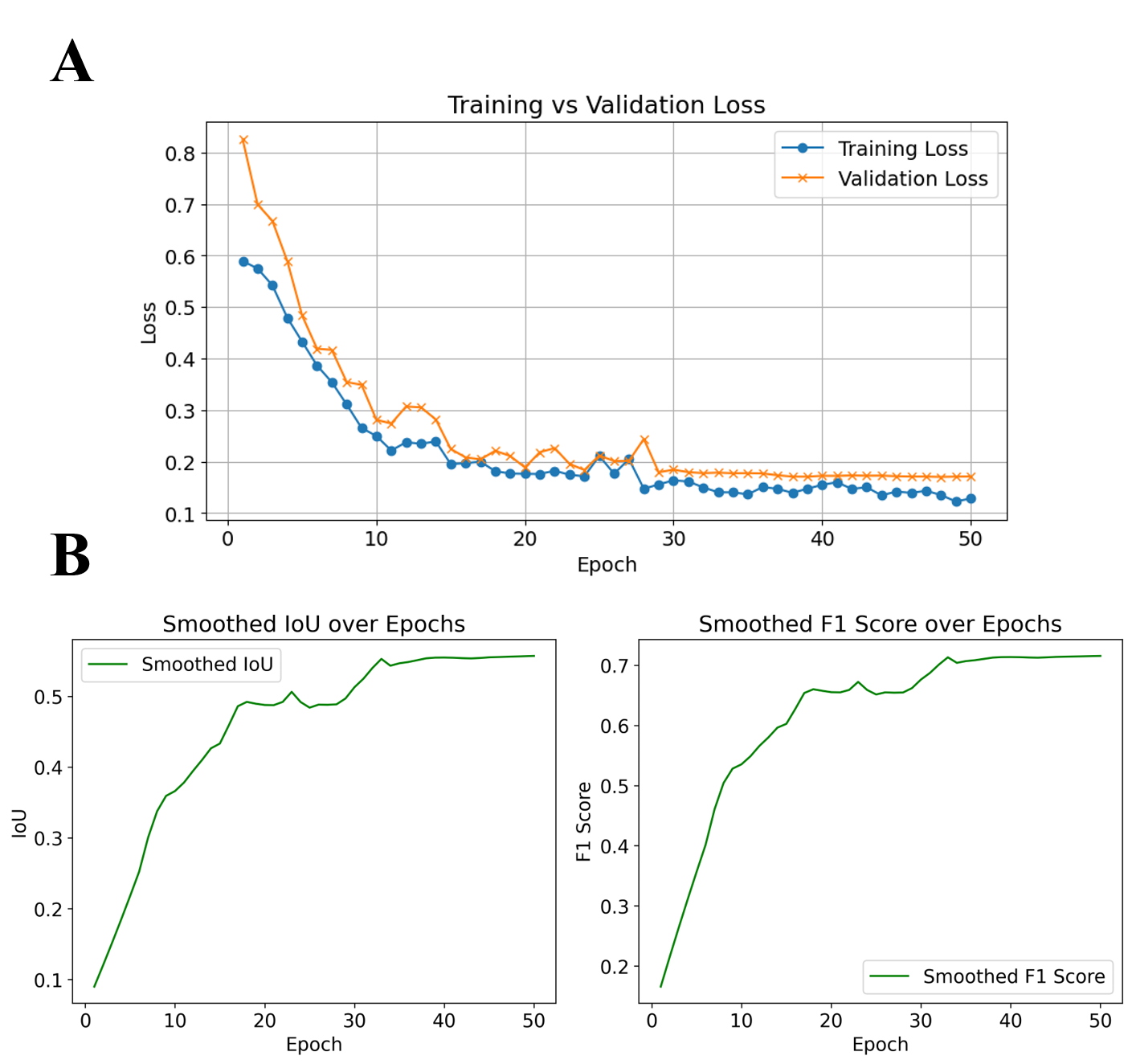}
    \caption{The training dynamics of \textit{CRISP} on the EMPIAR-10081 dataset. \textbf{A}: Training and validation loss versus epochs. \textbf{B}: IoU score versus epochs. \textbf{C}: F1 score versus epochs. Note that the Savitzky–Golay filter—a local polynomial regression method—is applied to smooth the noisy performance metrics across epochs by fitting a second-order polynomial over an 11-point moving window, effectively reducing high-frequency noise while preserving key signal features.}
    \label{fig:10081_metric}
\end{figure}

\begin{figure}[h]
    \centering
    \includegraphics[width=0.9\columnwidth]{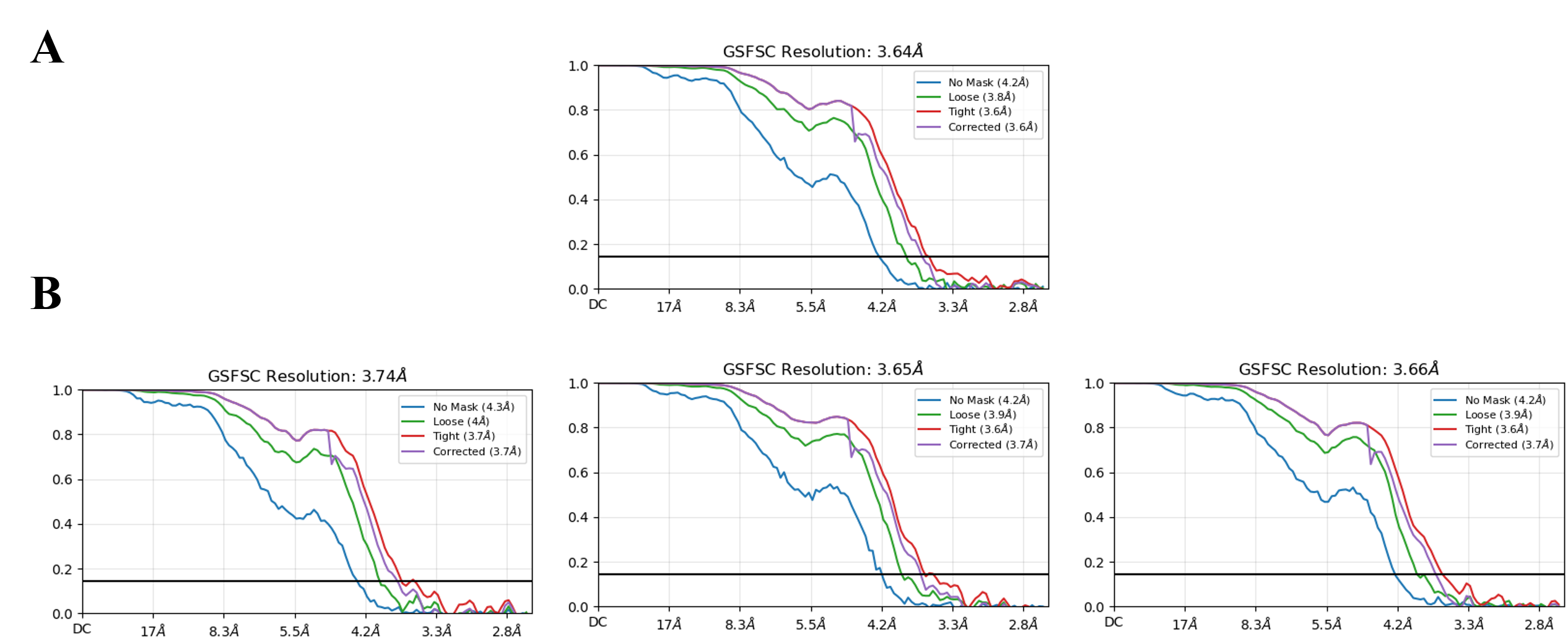}
    \caption{FSC curves obtained from 3D reconstruction with non-uniform refinement using particles extracted by different methods on the EMPIAR-10081 dataset. \textbf{A}: The FSC curve for CryoPPP. \textbf{B}: From left to right, FSC curves obtained using different center-finding algorithms: morphology and contour finding, the Crocker-Grier algorithm, and Non-Maximum Suppression.}
    \label{fig:10081_fsc_nonuniform}
\end{figure}

\end{appendices}

\end{document}